\documentclass[a4paper,final]{llncs}

\makeatletter
\renewcommand*{\thetable}{\arabic{table}}
\renewcommand*{\thefigure}{\arabic{figure}}
\let\c@table\c@figure
\makeatother

\newif\iflong
\longtrue

\iflong
\usepackage{longtable}
\fi

\usepackage{times}

\usepackage[T1]{fontenc}
\usepackage[scaled=0.81]{beramono}
\usepackage{hyperref}

\usepackage[pdftex]{graphicx}
\usepackage{amsmath,amssymb}
\usepackage{array,multirow}
\usepackage{xcolor}
\usepackage{xspace}
\usepackage{ifthen}

\usepackage{listings}
\usepackage{boogie}
\usepackage{why3lang}
\usepackage[T1]{fontenc}
\usepackage[scaled=0.81]{beramono} 
\newcommand{\tr}{\mathcal{T}}
\newcommand{\enc}{\mathcal{E}}
\newcommand{\des}{\mathcal{D}}
\DeclareRobustCommand{\openCurl}{\ensuremath{\text{\ttfamily\{}}}
\DeclareRobustCommand{\closeCurl}{\ensuremath{\text{\ttfamily\}}}}

\newcommand{\nat}{\textsc{nat}\xspace} 
\newcommand{\tes}{\textsc{tes}\xspace} 
\newcommand{\obj}{\textsc{obj}\xspace} 

\usepackage{tikz}
\usetikzlibrary{arrows,automata,shapes.symbols,positioning,calc}

\newcommand{\btw}{\texttt{b2w}\xspace}
\newcommand{\Boogie}{Boogie\xspace}
\newcommand{\WhyML}{WhyML\xspace}
\newcommand{\Why}{Why3\xspace}

\iflong
\newcommand{\tightParagraph}[1]{\paragraph{#1}}
\else
\makeatletter
\newcommand\tightParagraph{\@startsection{paragraph}{4}{\z@}%
                       {-5\p@ \@plus -4\p@ \@minus -4\p@}%
                       {-0.5em \@plus -0.22em \@minus -0.1em}%
                       {\normalfont\normalsize\itshape}}
\makeatother
\fi

\newcommand{\feature}[1]{\subsection{#1}}
\newcommand{\kw}[1]{\emph{#1}}

\newcommand{\typesOf}[1]{\mathit{types}(\MB{#1})}
\newcommand{\ctypesOf}[1]{\mathit{conc}(\MB{#1})}
\newcommand{\concPlus}[1]{\mathit{conc}^+(\MB{#1})}
\newcommand{\BBT}{\mathbb{T}}

\usepackage{changepage}

\begin{document}

\title{Why Just Boogie?}
\subtitle{Translating Between Intermediate Verification Languages}

\author{Michael Ameri\inst{1} \and Carlo A.\ Furia\inst{2}\thanks{Work done mainly while affiliated with ETH Zurich.}}

\institute{
Chair of Software Engineering, Department of Computer Science, \\
ETH Zurich, Switzerland $\quad$ \email{mameri@student.ethz.ch}
\and
Department of Computer Science and Engineering, \\
Chalmers University of Technology, Sweden $\quad$ \email{furia@chalmers.se}
}

\maketitle

\begin{abstract}
The verification systems \Boogie and \Why use their respective intermediate languages 
to generate verification conditions from high-level programs.
Since the two systems support different back-end provers (such as Z3 and Alt-Ergo) and are used to encode different high-level languages (such as C\# and Java), being able to translate between their intermediate languages would provide a way to reuse one system's features to verify programs meant for the other.
This paper describes a translation of \Boogie into \WhyML (\Why's intermediate language) that preserves semantics, verifiability, and program structure to a large degree.
We implemented the translation as a tool and applied it to 194 \Boogie-verified programs of various sources and sizes; \Why verified 83\% of the translated programs with the same outcome as \Boogie.
These results indicate that the translation is often effective and practically applicable.
\end{abstract}

\section{Introduction}\label{sec:intro}

Intermediate verification languages (IVLs) are intermediate representations used in verification technology.
Just like compiler design has benefited from decoupling front-end and back-end, IVLs help write verifiers that are more modular: the front-end specializes in encoding the rich semantics of a high-level language (say, an object-oriented language such as C\#) as a program in the IVL; the back-end generates verification conditions (VCs) from IVL programs in a form that caters to the peculiarities of a specific theorem prover (such as an SMT solver).

\Boogie~\cite{boogie-baseref} and \WhyML~\cite{FilliatreP13} are prime examples of popular IVLs with different, often complementary, features and supporting systems (respectively called \Boogie and \Why).
In this paper we describe a translation of \Boogie programs into \WhyML programs and its implementation as the tool \btw.
As we illustrate with examples in \autoref{sec:examples}, using \btw increases the versatility brought by IVLs: without having to design and implement a direct encoding into \WhyML\iflong or even being familiar with its peculiarities\fi, users can take advantage of some of the best features of \Why when working with high-level languages that translate to \Boogie. 

\tightParagraph{\Boogie vs.\ \WhyML.}
While the roles of \Boogie and \WhyML as IVLs are similar, the two languages have different characteristics that reflect a focus on complementary challenges in automated verification.
\Boogie is the more popular language in terms of front-ends that use it as IVL, which makes a translation \emph{from} \Boogie more practically useful than one into it; it has a finely tuned integration with the Z3 prover that results from the two tools having been developed by the same group (Microsoft Research's RiSE); it combines a simple imperative language with an expressive typed logic, which is especially handy for encoding object-oriented or, more generally, heap-based imperative languages.
In contrast, \WhyML has a more flexible support for multiple back-end provers it translates to, including a variety of SMT solvers as well as interactive provers such as Coq; it can split VCs into independent goals and dispatch them to different provers; if offers limited imperative constructs within a functional language that belongs to the ML family, which brings the side benefit of being able to \emph{execute} \WhyML programs---a feature quite useful to debug and validate verification attempts.

\tightParagraph{Goals and evaluation.}
The overall goal of this paper is devising a translation $\tr$ from \Boogie to \WhyML programs.
\iflong
The translation, described in \autoref{sec:translation}, should preserve correctness, verifiability, and readability as much as possible.
\fi
Preserving correctness means that, given a \Boogie program $p$, if its translation $\tr(p)$ is a correct \WhyML program then $p$ is correct (soundness); the converse should also hold as much as possible: if $\tr(p)$ is incorrect then $p$ is too (precision).
Preserving verifiability means that, given a \Boogie program $p$ that verifies in \Boogie, its translation $\tr(p)$ is a \WhyML program that verifies in \Why.
\iflong
Preserving readability means that the translation should not introduce unnecessary changes in the structure of programs.
\fi

The differences, outlined above, between \Boogie and \WhyML and their supporting systems make achieving correctness\iflong,\fi\ verifiability\iflong, and readability\fi\ challenging.
While we devised $\tr$ to cover the entire \Boogie language, its current implementation \btw does not fully support a limited number of features (branching, the most complex polymorphic features, and bitvectors) that make it hard to achieve verifiability in practice.
%
In fact, while replacing branching (goto) with looping is always possible~\cite{Harel-folk}, a general translation scheme does not produce verifiable loops since one should also infer invariants (which are often cumbersome due to the transformation).
Polymorphic maps are supported to the extent that their type parameters can be instantiated with concrete types; this is necessary since \WhyML's parametric polymorphism cannot directly express all usages in \Boogie, but it may also introduce a combinatorial explosion in the translation; hence, \btw fails on the most complex instances that would be unmanageable in \Why.
\Boogie's bitvector support is much more flexible than what provided by \Why's libraries; hence \btw may render the semantics of bitvector operations incorrectly.

These current implementation limitations notwithstanding (see \autoref{sec:translation} for details), we experimentally demonstrate that \btw is applicable and useful in practice.
As \autoref{sec:experiments} discusses, we applied \btw to 194 \Boogie programs of different size and sources; most of the programs have not been written by us and exercise \Boogie in a variety of different ways.
For 83\% (161) of these programs, \btw produces a \WhyML translation that \Why can verify as well as \Boogie can verify the original, thus showing the feasibility of automating translation between IVLs. 

\tightParagraph{Tool availability.}
The tool \btw is available as open source at:
\iflong\begin{center} \fi
\url{https://bitbucket.org/michael_ameri/b2w/}
\iflong\end{center}\fi

\section{Related Work}\label{sec:related-work}

\tightParagraph{Translations and abstraction levels.}
Translation is a ubiquitous technique in computer science; however, the most common translation schemes bridge \emph{different abstraction levels}, typically encoding a program written in a high-level language (such as Java) into a lower-level representation which is suitable for execution (such as byte or machine code).
Reverse-engineering goes the opposite direction---from lower to higher level---for example to extract modular and structural information from C programs and encode it using object-oriented constructs~\cite{TFNM-ECOOP13}.
This paper describes a translation between intermediate languages---\Boogie and \Why---which belong to \emph{similar abstraction levels}.
In the context of model transformations~\cite{MensG06}, so-called bidirectional transformations~\cite{Stevens07} also target lossless transformations between notations at the same level of abstraction.

\tightParagraph{Intermediate verification languages.}
The Spec\# project~\cite{BarnettFLMSV11} introduced \Boogie to add flex\-i\-bil\-i\-ty to the translation between an object-oriented language \iflong(a dialect of C\#)\fi{} and the verification con\-di\-tions{}\iflong in the logic fragments supported by SMT solvers\fi.
\iflong
An intermediate verification language embodies the idea of intermediate representation---a technique widespread in compiler construction---in the context of verification.
\fi{} 
Since its introduction for Spec\#, \Boogie has been adopted as intermediate verification language for numerous other front-ends such as 
Dafny~\cite{Leino04}, AutoProof~\cite{TFNP-TACAS15}, Viper~\cite{HeuleKMS13}, and Joogie~\cite{ArltS12}; 
its popularity demonstrates the advantages of using intermediate verification languages.

While \Boogie retains some support for different back-end SMT solvers, Z3\iflong~\cite{Z3}\fi{} remains its\iflong{} fully supported \fi{} primary target.
By contrast, supporting multiple, different back-ends is one of the main design goals behind the \Why system~\cite{FilliatreP13}\iflong, which does not merely generate verification conditions in different formats but offers techniques to split them into independently verifiable units and to dispatch each unit to a different prover\fi.
\Why also fully supports interactive provers,\iflong\footnote{In comparison, \Boogie's support for HOL is restricted and not up-to-date~\cite{BohmeLW08}.}\fi{} which provide a powerful means of discharging the most complex verification conditions that defy complete automation.

\iflong
Another element that differentiates \Boogie and \Why is the support for executing programs; this is quite useful for debugging verification attempts and for applying testing-like techniques to the realm of verification.
Boogaloo~\cite{PFW-RV13} supports symbolic execution of \Boogie programs; Symbooglix is a more recent project with the same goal~\cite{symbooglix}. 
Thanks to it being a member of the ML family, \Why directly supports symbolic execution as well as compilation of \WhyML programs to OCaml.
\fi

In all, while the \Boogie and \WhyML languages belong to a similar abstraction level, they are part of systems with complementary features, which motivates this paper's idea of translating one language into the other.
\iflong
Since \Boogie is overall more popular, in terms of tools that use it as a back-end, the translation from \Boogie to \WhyML is more practically useful than the one in the opposite direction.
\fi

Other intermediate languages for verification are Pilar~\cite{Pilar-VSTTE}, used in the Sireum framework for SPARK; Silver~\cite{HeuleKMS13}, an intermediate language with native support for permissions in the style of separation logic; and the flavor of dynamic logic for object-oriented languages~\cite{SchmittUW10} used in the KeY system.
Another approach to generalizing and reusing different translations uses notions from model transformations to provide validated mappings for different high-level languages~\cite{ChengMP15}.
Future work may consider supporting some of these intermediate languages and approaches.

\section{Motivating Examples}\label{sec:examples}

Verification technology has made great strides in the last decade or two, but a few dark corners remain where automated reasoning shows its practical limitations.
\autoref{fig:motivating-examples} provides three examples of simple \Boogie programs that trigger incorrect or otherwise unsatisfactory behavior. We argue that translating these programs to \WhyML makes it possible to verify them using a different, somewhat complementary verification tool; overall, confidence in the results of verification is improved.

Procedure \B{not_verify} in~\autoref{fig:motivating-examples} has a contradictory postcondition (notice \B{N < N}, \B{N} is a nonnegative constant, and the loop immediately terminates).
Nonetheless, recent versions of \Boogie and Z3 
successfully verify it.\footnote{\url{https://github.com/boogie-org/boogie/issues/25}}
More generally, unless the complete toolchain has been formally verified (a monumental effort that has only been performed in few case studies~\cite{Leroy09,KleinAEHCDEEKNSTW10,KumarMNO14}), there is the need to \emph{validate} the successful runs of a verifier.
Translating \Boogie to \Why provides an effective validation, since \Why has been developed independent of \Boogie and uses a variety of backends that \Boogie does not support.
Procedure \B{not_verify} translated to \Why (\autoref{fig:translated-motivating-examples}) does not verify as it should.

Procedures \B{lemma_yes} and \B{lemma_no} in \autoref{fig:motivating-examples} demonstrate \Boogie's support for mathematical real numbers, which is limited in the way the power operator \B{**} is handled. \Boogie vacuously verifies both properties $2^3 > 0$ and $2^3 < 0$, even though Z3 outputs some unfiltered errors that suggest the verification is spurious{}\iflong{} (the power operator \B{**} is not properly supported); indeed, only the inequality encoded by \B{lemma_yes} is correct\fi.
\Why provides a more thorough support for real arithmetic\iflong, both by translating to backends such as Alt-Ergo and by providing a more effective encoding in Z3\fi; 
in fact, it verifies the translated procedure \B{lemma_yes} but correctly fails to verify \B{lemma_no}.

The loop in procedure \B{trivial_inv} in \autoref{fig:motivating-examples} includes an invariant asserting that \B{i} takes only even values.
Even if this is clearly true, \Boogie fails to check it; 
pinning down the precise cause of this shortcoming requires knowledge of \Boogie's (and Z3's) internals, although it likely is a manifestation of the ``triggers'' heuristics that handle  (generally undecidable) quantified expressions. 
\iflong
Based on this knowledge, there are specification patterns that try to work around such idiosyncrasies; in the example, one could introduce a ``witness'' ghost variable \B{k} such that \B{i == 2*k} is an invariant.
\fi
However, if we insist on verifying the program in its original form, \Why can dispatch verification conditions to \emph{interactive} provers, where the user provides the crucial proof steps.\footnote{\Why can also check the invariant automatically by relying on the CVC4 SMT solver.}
Cases such as the loop invariant of \B{trivial_inv} where a proof is ``obvious'' to a human user but it clashes against the default strategies to handle quantifiers are prime candidate to exploit interactive provers.
\iflong
Thus, translating \Boogie to \Why provides another means of exploiting the latter's versatile support for interactive provers and multiple backends.
\fi

\begin{figure}[!thb]
\centering
\scriptsize
\lstset{language=boogie}
\begin{tabular}{p{.33\textwidth} p{.3\textwidth} p{.3\textwidth}}
{\begin{lstlisting}
const N: int;
axiom 0 <= N;

procedure not_verify()
  ensures (forall k, l: int :: 
   0 <= k <= l < N ==> N < N);
{
  var x: int;
  x := -N;
  while (x != x) { }
}\end{lstlisting}}
&
{\begin{lstlisting}
procedure lemma_yes()
  ensures 2.0**3.0 > 0.0;
{  }

procedure lemma_no()
  ensures 2.0**3.0 < 0.0;
{  }\end{lstlisting}}
&
{\begin{lstlisting}
procedure trivial_inv()
{
  var i: int;
  i := 0;
  while (i < 10)
   invariant 0 <= i <= 10;
   invariant 
    (exists j: int :: i == 2*j);
  { i := i $\;$+ 2; }
}\end{lstlisting}}
\end{tabular}
\caption{Three simple \Boogie programs for which automated reasoning is limited.}
\label{fig:motivating-examples}
\end{figure}

\section{Boogie-to-Why3 Translation}\label{sec:translation}

Intermediate languages for verification combine programming constructs and a logic language.
When used to encode programs written in a high-level language, the programming constructs encode program behavior, and the logic constructs encode specifications, constrain the semantics to conform to the high-level language's (typically through axioms), and support other kinds of annotations (such as triggers).

Both \Boogie and \WhyML provide, as logic language, a typed first-order logic with arithmetic.
\Boogie's programming constructs are a simple imperative language with both structured (while loops, procedures) and unstructured (jumps, global variables) statements.
\WhyML's programming constructs combine an ML-like functional language with a few structured imperative features such as mutable variables and loops.

Correspondingly, we define a translation $\tr \colon \textsl{\Boogie} \to \textsl{\WhyML}$ of \Boogie to \WhyML as the composition $\enc{} \circ \des{}$ of two translations: $\des \colon \textsl{\Boogie} \to \textsl{\Boogie}$ is a desugaring\footnote{This is unrelated to \Boogie's built-in desugaring mechanism (option \B{/printDesugared}).} which rewrites away the \Boogie constructs, such as \emph{call-forall}, that have no similar construct in \WhyML by expressing them using other features of \Boogie.
Then, $\enc \colon \textsl{\Boogie} \to \textsl{\WhyML}$ encodes \Boogie programs simplified by $\des$ as \WhyML programs\iflong, while introducing constraints that ensure that the semantics in \WhyML mirrors the one in \Boogie\fi.
For simplicity, the presentation does not sharply separate the two translations $\des{}$ and $\enc{}$ but defines either or both of them as needed to describe the translation of arbitrary \Boogie constructs.

A single feature of the \Boogie language significantly compounds the complexity of the translation: \emph{polymorphic maps}\iflong, which correspond to mappings between domains of generic type\fi.
\iflong
\Why does support polymorphic maps through a library, but its type system is more restrictive and does not allow the same degree of freedom as \Boogie's in using variables of polymorphic map types.
\fi
For clarity, the presentation of the translation initially ignores polymorphic maps.
Then, \autoref{sec:poly-maps-all} discusses how the general translation scheme can be extended to support them.

As running examples, \autoref{fig:translated-motivating-examples} shows how $\tr$ translates the examples of \autoref{fig:motivating-examples}. 

\begin{figure}[!tb]
\centering
\scriptsize
\lstset{language=why3,xleftmargin=0pt}
\begin{tabular}{p{.3\textwidth} p{.31\textwidth} p{.31\textwidth}}
{\begin{lstlisting}
constant N: int
axiom A0: 0 <= N;

val not_verify (): ()
  ensures { forall k, l: int .
   0 <= k <= l < N -> N < N }

let not_verify_impl(): ()
  ensures { forall k, l: int .
   0 <= k <= l < N -> N < N }
 =(
   let x = ref (any int) in
    x.contents <- -N;
    while 
     (x.contents <> x.contents)
    do done;
  end )\end{lstlisting}}
&
{\begin{lstlisting}
val lemma_yes (): ()
  ensures
    { (pow 2.0 3.0) >. 0.0 }

val lemma_no (): ()
  ensures
    { (pow 2.0 3.0) <. 0.0 }

let lemma_yes_impl (): ()
  ensures
    { (pow 2.0 3.0) >. 0.0 }
=( )

let lemma_no_impl (): ()
  ensures
    { (pow 2.0 3.0) <. 0.0 }
=( )\end{lstlisting}}
&
{\begin{lstlisting}
val trivial_inv (): ()

let trivial_inv_impl (): ()
=(
  let i = ref (any int) in
   i.contents <- 0;
   while (i.contents < 10) do
    invariant
      { 0 <= i.contents <= 10 }
    invariant
      { exists j: int . 
          i.contents = 2*j }
    i.contents <- i.contents + 2;
   done;
 )\end{lstlisting}}
\end{tabular}
\caption{The translation to \WhyML of the three \Boogie programs in \autoref{fig:motivating-examples}. \iflong(Boilerplate such as general declarations, imports, and frame condition checking are omitted for clarity.)\fi}
\label{fig:translated-motivating-examples}
\end{figure}

\feature{Types} \label{sec:types}
\iflong
\Boogie types include primitive types, instantiated type constructors, and map types.
\fi

\tightParagraph{Primitive types}
are \B{int} (mathematical integers), \B{real} (mathematical reals), {}\B{bool} (Bool\-eans)\iflong, and \B{bv$n$} ($n$-bit vectors)\fi.
$\tr$ translates primitive types into their \Why analogues as shown in \autoref{tab:primitive-types}.
\iflong
Since \Why offers primitive types and operations on them through libraries, $\tr$ also generates import statements for the libraries that provide the same operations that are available in \Boogie, such as integer to/from real conversion. 
\fi

\begin{table}
\centering
\footnotesize
\setlength{\tabcolsep}{3pt}
\begin{tabular}{l l r}
$T$ & $\tr(T)$ & \Why libraries \\
\hline 
{\B{int}} & {\W{int}} & {\W{int.Int}}, {\W{int.EuclideanDivision}} \\
{\B{real}} & {\W{real}} & {\W{real.RealInfix}}, {\W{real.FromInt}}, {\W{real.Truncate}}, {\W{real.PowerReal}} \\
{\B{bool}} & {\W{bool}} & {\W{bool.Bool}} \\
\iflong {\B{bv$n$}} & {\W{bv}} & {\W{bv.BitVector with constant size = $\:n$}} \fi
\end{tabular}
\caption{Translation of primitive types, and \Why libraries supplying the necessary operations.}
\label{tab:primitive-types}
\end{table}

\tightParagraph{Type constructors.}
A \Boogie type declaration using the \kw{type constructor} syntax \iflong\footnote{$\tr$ ignores the optional type  modifier \B{finite}, since it does not seem fully supported in Boogie.}\fi{} introduces a new parametric type \B{T}\iflong with parameters \B{a$_1$}, $\ldots$, \B{a$_m$}\fi.
$\tr$ translates it to an algebraic type with constructor~\W{T}:
{$\tr($ \B{type T a$_1$ $\ldots$ a$_m$} $)\ =\ $ \W{type T 'a$_1$ $\ldots$'a$_m$}}
for $m \geq 0$, where ticks \W{'} identify type parameters in \WhyML.


\tightParagraph{Map types.}
A \Boogie \kw{map type} \B{M} declared as:
$\MB{type M = [T}_1\MW{,} \ldots \MB{T}_n\MB{] U}$
defines the type of a mapping from $\MB{T}_1 \times \cdots \times \MB{T}_n$ to $\MB{U}$, for $n \geq 1$.
\Why supports maps through its library \W{map.Map};{}\iflong\footnote{\Why's maps, like \Boogie's, do not satisfy extensionality~(\url{http://lists.gforge.inria.fr/pipermail/why3-club/2013-February/000572.html}).}\fi{} 
hence, $\tr(\MB{M}) = \MW{map (}\tr(\MB{T}_1)\MW{,} \ldots\MW{,} \tr(\MB{T}_n)\MW{)}\:\tr(\MB{U})$, where an $n$-tuple encapsulates the $n$-type domain of \B{M}.

\feature{Constants}

The translation of constant declarations is generally straightforward, following the scheme:
\begin{center}
$\tr ($\B{const c: T}$) \ =\ $ \W{constant c: $\;\tr($T$)$}
\end{center}

\iflong
\tightParagraph{Unique constants.}
All constants of a type \B{T} declared with the modifier \B{unique} have values that are pairwise different.
Thus, for $m$ constants \B{const unique c$_1$, $\ldots$, c$_m$: T}, $\tr$ encodes the uniqueness properties using $m \choose 2$ axioms\iflong $\MW{axiom unique_c_i_j: c}_i \;\MW{ <> c}_j$, for $1 \leq i \neq j \leq m$\fi.

\tightParagraph{Orders.}
\Boogie provides the operator \B{<:} to express partial order over every type; $\tr$ introduces a polymorphic operator \W{<:} and axiomatizes its reflexive, antisymmetric, and transitive properties\iflong:\else; it also axiomatizes \Boogie's partial-order relation constraints.\fi

\iflong
\begin{why3}[numbers=none]
predicate (<:) (x: 'a) (y: 'a)
axiom ReflexivePO:     forall x: 'a     . x <: x
axiom AntisymmetricPO: forall x y: 'a   . x <: y && y <: x -> x = y
axiom TransitivePO:    forall x y z: 'a . x <: y && y <: z -> x <: z
\end{why3}
\fi

\iflong
\Boogie supplies special syntax to describe a partial-order relations with a certain structure, which corresponds to a DAG where any two nodes $x$ and $y$ are connected by an edge $x \to y$ iff $x \:\MB{<:}\: y$ and $y$ is a direct successor of $x$ in the order.
Let \B{a}, \B{b}, \B{c}, \B{d}, \B{e}, \B{f} be unique\footnote{Uniqueness is not required but makes the order specification easier to present.} constants of the same type \B{T}.
The \Boogie syntax to specify ordering between them is in \autoref{fig:ordering-specs}.
$\des$ reconstructs the DAG of the order specification, and then formalizes it in axiomatic form.
For example, the specifications in \autoref{fig:ordering-specs} determine the DAG in \autoref{fig:DAG-specification}, which is axiomatized as in \autoref{fig:axioms-of-DAG}.

\begin{figure}
\centering
\begin{tabular}{l c@{$\ $} p{0.6\textwidth}}
\multicolumn{1}{c}{\textsc{\Boogie specification}}  &&  \multicolumn{1}{c}{\textsc{Semantics}} \\
\hline
{\B{const c: T extends a, b;}} && {\B{a}} and {\B{b}} are the only direct successors of {\B{c}} \\
{\B{const a: T extends;}}      && {\B{a}} has no (direct) successors \\
{\B{const d: T extends c complete;}}  && {\B{c}} has no direct predecessors other than {\B{d}} and any others that are explicitly specified \\
{\B{const e, f: T extends unique d;}}  && {\B{d}} is the only direct successor of both {\B{e}} and {\B{f}}, and the subgraphs that originate in {\B{e}} and {\B{f}} are disjoint
\end{tabular}
\caption{Ordering specifications in \Boogie (older versions of \Boogie use \B{<:} instead of \B{extends}).}
\label{fig:ordering-specs}
\end{figure}

\begin{figure}
\centering
\begin{tikzpicture}[->]
\matrix[row sep=0pt, column sep=7mm] {
\node (lb) {}; & \node (b) {\B{b}}; & \node(rb) {};     &                   & \node (e) {\B{e}}; & \node(re) {}; \\
\node (la) {}; & \node (a) {\B{a}}; & \node(c) {\B{c}}; & \node(d) {\B{d}}; &                    &               & \node (bottom) {}; \\
               &                    & \node(uc) {};     &                   & \node (f) {\B{f}}; & \node(urf) {}; \\
};
\path (c) edge (b);
\path (c) edge (a);
\path (d) edge (c);
\path (e) edge (d);
\path (f) edge (d);
\begin{scope}[dotted]
\path (b) edge (lb);
\path (b) edge (la);
\path (rb) edge (b);
\path (uc) edge (a);
\path (re) edge (e);
\path (urf) edge (f);
\end{scope}
\path (bottom) edge[dashed,-] (d);
\end{tikzpicture}
\caption{DAG corresponding to the ordering specification of \autoref{fig:ordering-specs}. Solid edges denote the successor relation; dotted edges denote allowed (but not specified) relations; the dashed line expresses disjointness of the two sub-graphs.}
\label{fig:DAG-specification}
\end{figure}
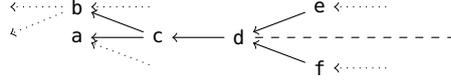

\begin{figure}
\begin{boogie}[numbers=none,basicstyle=\scriptsize\ttfamily,xleftmargin=20mm]
axiom (c <: a && c <: b && forall $x$: T :: $\:$c <: $\:x$ ==> c == $\;x$ || a <: $\;x$ || b <: $\;x$)
axiom (forall $x$: T :: $\ $!(a <: $\:x$))
axiom (d <: c && forall $x$: T :: $\:$$x$ <: $\:$c ==> c == $\:x$ || $x$ <: $\:$d)
axiom (e <: d && forall $x$: T :: $\:$e <: $\:x$ ==> e == $\:x$ || d <: $\;x$)
axiom (f <: d && forall $x$: T :: $\:$f <: $\:x$ ==> f == $\:x$ || d <: $\;x$)
axiom (forall $x$: T :: $\:x$ <: $\ $e ==> $\ $!($\:x$ <: $\;$f))
axiom (forall $x$: T :: $\:x$ <: $\ $f ==> $\ $!($\:x$ <: $\;$e))

\end{boogie}
\caption{Axiomatization of the ordering specification in \autoref{fig:ordering-specs}.}
\label{fig:axioms-of-DAG}
\end{figure}
\fi
\fi

\feature{Variables} \label{sec:variables}
\Why supports mutable variables through the reference type \W{ref} from theory \B{Ref}.
Boogie global variable declarations become global value declarations of type \W{ref}; Boogie local variable declarations become \W{let} bindings with local scope.
Thus, if \B{v} is a global variable and \B{l_v} is a local variable in \Boogie:
\begin{center}
\begin{tabular}{lc@{$\ $}rl}
global variable && {$\tr($\B{var v: T}$)$} & {$=\ $\W{val v: ref }$\;\tr($\W{T}$)$}  \\
local  variable && {$\tr($\B{var l_v: T}$)$} & {$=\ $\W{let l_v = ref (any }$\;\tr($\W{T}$)$\W{) in }} 
\end{tabular}
\end{center}
The expression \W{any T} provides a nondeterministic value of type \W{T}.

\feature{Functions}
\Boogie function \emph{declarations} become \WhyML function declarations:
\begin{multline}
{\tr\big(\MB{function f(x}_1\MB{: T}_1\MB{, }\ldots\MB{, x}_n\MB{: T}_n\MB{) returns (U)}\big)}
\\=\ 
{\MW{function f (x}_1\MW{:}\tr(\MW{T}_1)\MW{)}\cdots\MW{(x}_n\MW{:}\tr(\MW{T}_n)\MW{): }\tr(\MW{U})}
\label{eq:tr-function-decl}
\end{multline}

\noindent
\WhyML function \emph{definitions} require, unlike \Boogie's, a variant to ensure that recursion is well-formed.
Therefore, \Boogie function definitions are not translated into \WhyML function definitions but are axiomatized\iflong: if function \B{f} in \eqref{eq:tr-function-decl} has body \B{B}, $\des$ replaces the body with the $\MB{axiom (forall z}_1\MB{: T}_1\MB{,}\ldots\MB{,}\MB{z}_n\MB{: T}_n\MB{:: f(z}_1\MB{,}\ldots\MB{,}\MB{z}_n\MB{) = B)}$\fi.\footnote{To take advantage of \Why's well-formedness checks, we plan to offer translations of \Boogie functions to \WhyML functions as a user option in future work.}

\feature{Expressions} \label{sec:expressions}
\iflong
The translation of \Boogie expressions to \WhyML expressions is mostly straightforward, given the translation of types described above.
We describe the few cases that deserve some detail.
\fi

\iflong
\tightParagraph{Nondeterministic choice.}
The special value \B{*} represent a nondeterministic Boolean choice (used in loop exit flags and conditionals); we define $\tr(\MW{*}) = \MW{any bool}$, which provides a nondeterministic Boolean value. \fi

\tightParagraph{Variables.}
Since a \Boogie variable \B{v} of type \B{T} turns into a value \W{v} of type \W{ref$\:\tr($T$)$}, occurrences of \B{v} in an expression translate to \W{v.contents}, which represents the value attached to reference \W{v}.

\tightParagraph{Map expressions.}
$\tr$ translates map selection and update using functions \W{get} and \W{set} from theory \W{Map}.
If \B{m} is a map of type \B{M} defined in \autoref{sec:types}, then:\iflong\footnote{Despite its name, \W{set} returns a new map rather than changing its argument's value.}\fi{} 
\begin{center}
\setlength{\tabcolsep}{6pt}
\begin{tabular}{lll}
&  \multicolumn{1}{c}{$E$} & \multicolumn{1}{c}{$\tr(E)$} \\
\hline
selection 
  & {\B{m[e$_1$, $\ldots$, e$_n$]}}
  & {\W{get}$\;\tr($\B{m}$)\;$\W{(}$\tr($\B{e}$_1)$\W{,}$\ldots$\W{,}$\tr($\B{e}$_n)$\W{)}} \\
update 
  & {\B{m[e$_1$, $\ldots$, e$_n$ := $\;$f]}}
  & {\W{set}$\;\tr($\B{m}$)\;$\W{(}$\tr($\B{e}$_1)$\W{,}$\ldots$\W{,}$\tr($\B{e}$_n)$\W{)}$\:\tr($\W{f}$)$}
\end{tabular}
\end{center}

\tightParagraph{Lambda expressions.}
\iflong \Boogie recently introduced lambda expressions as syntactic sugar for maps. 
While \WhyML has lambda abstractions, they are not allowed as first-order values in programs~\cite{ClochardFMP14}.
Instead, the \fi{} translation desugars lambda expression into constant maps:
$\des(\MB{lambda x}_1\MB{: T}_1\MB{,}\ldots\MB{,x}_n\MB{: T}_n\MB{ :: e}) = \MB{lmb}$, where $\MB{const lmb}: [\MB{T}_1\MB{,}\ldots\MB{,T}_n] \tau(\MB{e})$ is axiomatized by
$\MB{axiom (forall x}_1\MB{: T}_1\MB{,}\ldots\MB{,x}_n\MB{: T}_n \MB{:: lmb[x}_1\MB{,}\ldots\MB{,x}_n\MB{] == e)}$, and $\tau(\MB{e})$ is \B{e}'s type.

\iflong
\tightParagraph{Old expression.}
Within a procedure's postcondition or body, the expression \B{old(e)} refers to the value of \B{e} in the prestate.
\WhyML offers a more general construct to refer to an expression's value at any labeled point within a procedure's body.
Hence, every \WhyML procedure implementation translating a \Boogie procedure implementation includes a label \W{"begin"}, so that $\tr(\MB{old(e)})$ is just $\MW{old}\ \tr(\MB{e})$ within postconditions, and is $\MW{at}\ \tr(\MB{e})\ \MB{'}\ \MW{"begin"}$ within bodies.
\fi

\iflong
\tightParagraph{Bitvectors.}
\Why's theory \W{BitVectors} does not provide all operations that are supported by \Boogie.
In particular, it does not support \emph{extraction expressions} \B{b[n:m]} (drop the \B{m} least significant bits and return the next $\MB{n} - \MB{m}$ least significant bits) and \emph{concatenation expressions} \B{b +$\;$+ c} (the bit vector obtained by concatenating \B{b} and \B{c}).
$\tr$ introduces functions \W{extract (b: bv) (n: int) (m: int): bv} and \W{cat (b: bv) (c: bv): bv} 
and uses them to translate applications of these bit vector operators, but leaves them uninterpreted in \Why.
$\tr$'s implementation currently supports only the bitvectors operations available in \Why's theory \W{BitVectors}.
\fi

\feature{Procedures}

\Boogie procedures have a declaration (signature and specification) and zero or more implementations.
The latter follow the general syntax of \autoref{fig:tr-procedures} (left)\iflong, where a procedure \B{p} with input argument \B{t} and output argument \B{u} has one implementation with local variable \B{l} and body \B{B}\fi.
For simplicity of presentation, \B{p} has one input argument, one output argument, and one local variable, but generalizing the description to an arbitrary number of variables is straightforward.

\begin{figure}[!htb]
\lstset{xleftmargin=5mm}
\scriptsize
\begin{tabular}{p{.4\textwidth} p{.45\textwidth}}
{\begin{lstlisting}[language=boogie]
procedure p(t: T where Wt) 
  returns (u: U where Wu);
  requires      R;
  free requires fR;
  modifies      M;
  ensures       E;
  free ensures  fE;

implementation p(t: T) 
  returns (u: U) 
  { 
    var l: L where Wl; 
    B 
  }\end{lstlisting}} 
&
\iflong
{\begin{lstlisting}[language=Why3]
val p (t : $\tr($T$)$): $\tr($U$)$
  requires { $\tr($R$)$ }
  writes   { M }
  returns  { | u -> $\tr($E$)$ }
  returns  { | u -> $\tr($fE$)$ }
  returns  { | u -> $\tr($Wu$)$ }

let p_impl0 (t: $\tr($T$)$): $\tr($U$)$
  requires { $\tr($R$)$ } requires { $\tr($fR$)$ }
  returns  { | u -> $\tr($E$)$ }
=(
  $\tr($$\text{\textcolor[HTML]{6699FF}{\bfseries var}}$ u: U; $\text{\textcolor[HTML]{6699FF}{\bfseries var}}$ l: L;$)$
  assume { $\tr($Wg$)$ }  -- where of globals
  assume { $\tr($Wt$)$ }  -- where of inputs
  assume { $\tr($Wl$)$ }  -- where of locals
  assume { $\tr($Wu$)$ }  -- where of outputs
  try ( $\tr($B$)$ )
  with | Return -> assume { true } end
  $\tr($u$)$
 )

let p_impl0_frame (t: $\tr($T$)$): $\tr($U$)$
  requires { $\tr($R$)$ } requires { $\tr($fR$)$ }
  writes { M }
  reads { $G$ } -- all globals
  returns  { | u -> true }
=( ... -- as in $\text{p\_impl0}$
   $\tr(m$ := $m)\textrm{, for }m\in$ M
   assume { yes($g$) }$\textrm{, for }g\in G$
   $\tr($u$)$ )\end{lstlisting}}
\fi
\end{tabular}
\caption{Translation of a \Boogie procedure (left) into \WhyML (right).}
\label{fig:tr-procedures}
\end{figure}

The specification of procedure \B{p} consists of preconditions \B{requires}, frame specification \B{modifies}, and postconditions \B{ensures}.
\iflong
A precondition is an assertion that callers of \B{p} must satisfy upon calling, and that every implementation of \B{p} can assume; \B{free} preconditions need not be satisfied by callers.
A postcondition is an assertion that every implementation of \B{p} must satisfy upon terminating, and that every caller of \B{p} can assume; \B{free} postconditions need not be satisfied by implementations.
Every implementation of \B{p} may only modify the global variables listed in \B{p}'s frame specification.
\fi

$\tr$ translates a generic procedure \B{p} as shown in \autoref{fig:tr-procedures} (right).
The declaration of \B{p} determines \W{val p}, which defines the semantics of \B{p} for clients: the \B{free} precondition \B{fR} does not feature there because clients don't have to satisfy it, whereas both \B{free} and non-\B{free} postconditions are encoded as \B{returns} conditions.
The implementation of \B{p} determines \W{let p_impl0}, which triggers the verification of the implementation against its specification: both \B{free} and non-\B{free} preconditions are encoded, whereas the \B{free} postcondition \B{fE} does not feature there because implementations don't have to satisfy it.
The body introduces \W{let} bindings for the local variable \W{l} and for a new local variable~\W{u} which represents the returned value; these declarations are translated as discussed in \autoref{sec:variables}.
Then, a series of \W{assume} encode the semantics of \Boogie's \B{where} clauses, which constrain the nondeterministic values variables can take (\B{Wg} comes from any global variables, which are visible everywhere); 
\B{p}'s body \B{B} is translated and wrapped inside an exception-handling block \W{try}, which does not do anything other than allowing abrupt termination of the body's execution upon throwing a \W{Return} exception (see \autoref{sec:statements} for details).
Regardless of whether the body terminates normally or exceptionally, the last computed value of \W{u} is returned in the last line, and checked against the postcondition in \W{returns}.
\iflong
Another implementation \W{let p_impl0_frame} checks the frame condition (\B{modifies} clause).\footnote{The tool \btw does not currently implement frame condition checks.}
It relies on the same full precondition as \W{p_impl0} but has postcondition \W{true} since \B{E} has already been checked; it includes a \W{writes} clause and a \W{reads} clause.
\Why checks that a global variable is in the \W{writes} clause if and only if it is written by the implementation; since \Boogie's \B{modifies} clause only expresses variables that \emph{may} be written, \W{p_impl0_frame} includes an assignment of every variable in \B{M} to itself so that the requirement that every variable in \W{M} is written is vacuously satisfied.
When a \W{writes} clause is present, \Why also requires a \W{reads} clause and checks that every variable in it is written, read, or both.
The translation builds a \W{reads} clause with all global variables $G$, and vacuously reads all of them using function \W{yes 'a: bool}, which identically returns \W{true} for any input; this makes the \B{reads} clause satisfied by any implementation.
\fi
In all, the modular semantics of \Boogie's procedure \B{p} is preserved.

\feature{Statements}
\label{sec:statements}

\iflong
\tightParagraph{Axioms and assertions.}
\Boogie's \B{assert e}, \B{assume e}, and \B{axiom e} statements translate to 
$\MW{assert \{}\,\tr(\MW{e})\,\MW{\}}$, 
$\MW{assume \{}\,\tr(\MW{e})\,\MW{\}}$, and 
$\MW{axiom A:}\,\tr(\MW{e})$
in \WhyML.
\fi

\tightParagraph{Assignments.}
Assignments involve variables (global or local), which become mutable references in \WhyML: 
$\tr(\MB{v := e}) = \MW{v.contents <-}\;\tr(\MW{e})$.
\Boogie parallel assignments become simple assignments using \W{let} bindings of limited scope:
\begin{equation}
{\tr(\MB{v}_1\MB{,}\ldots\MB{,}\MB{v}_m\MB{ := e}_1\MB{,}\ldots\MB{,}\MB{e}_m)}
\ =\ \left\{
\begin{tabular}{l}
{\W{let}$\;$\W{e'}$_1$\W{=}$\tr($\W{e}$_1)$\W{,}$\ldots$\W{,}\W{e'}$_m$\W{=}$\tr($\W{e}$_m)\;$\W{in}} \\
{$\quad $$\tr($\B{v}$_1\,$\B{:=}$\,$\W{e'}$_1$$)$\W{;}$\cdots$\W{;}$\tr($\B{v}$_m\,$\B{:=}$\,$\W{e'}$_m$$)$}
\end{tabular}\right.
\end{equation}

\tightParagraph{Havoc.}
An abstract function \W{val havoc (): 'a} provides a fresh, nondeterministic\iflong\footnote{\url{http://lists.gforge.inria.fr/pipermail/why3-club/2013-April/000615.html}}\fi{} value of any type \W{'a}.
It translates \Boogie's \B{havoc} statements following the scheme:\iflong\footnote{Alternatively, we could define $\tr(\MB{havoc v}) = \MW{any}\;\tr(\MB{T})$, where \B{T} is \B{v}'s type.}\fi{} 
\begin{equation*}
\footnotesize
\tr(\MB{havoc u, v}) \ =\ 
\tr(\MW{u})\MW{<-havoc();}\tr(\MW{v})\MW{<-havoc();}\MW{assume \{}\,\tr(\MB{Wu})\,\MW{\};assume \{}\,\tr(\MB{Wv})\,\MW{\}}
\end{equation*}
where \B{Wu} and \B{Wv} are the \B{where} clauses of \B{u}'s and \B{v}'s declarations; the generalization to an arbitrary number of variables is obvious.
It is important that the \W{assume} statements follow all the calls to \W{havoc}: since \B{Wv} may involve \B{u}'s value, \B{havoc u, v} is not in general equivalent to \B{havoc u; havoc v}; the translation reflects this behavior.

\tightParagraph{Return.}
The behavior of \Boogie's \B{return} statement, which determines the abrupt termination of a procedure's execution, is translated to \WhyML using exception handling.
An exception handling block wraps each procedure's body, as illustrated in \autoref{fig:tr-procedures}, and catches an \W{exception Return}; thus, $\tr(\MB{return}) = \MW{raise Return}$.

\iflong
\tightParagraph{Jumps (branching).}
In addition to structured \B{while} loops (discussed below), \Boogie provides jump statements of the form \B{goto l$_1$,$\ldots$,l$_n$}, which nondeterministically jump to any of the locations labeled by \B{l$_k$}.
The translation must remove jump statements in a way that preserves verifiability; this rules out ``global'' approaches using a program counter \cite{Harel-folk,TFNMO12-WCRE12}, since they would require new invariants about the counter.
Instead, we introduce simple heuristics that replace jumps with structured code; since the usage of jumps in \Boogie programs tend to follow well-defined patterns that can be traced back to structured loops, the heuristics may be sufficient in practice.\footnote{$\tr$'s implementation currently does not support this translation of \B{goto} statements.}

\begin{figure}[!hbt]
\begin{tikzpicture}[->,semithick, initial text=,
  state/.style={minimum width=15mm,minimum height=10mm,inner sep=1pt,align=center,draw,rectangle},
  node distance=7mm and 7mm,align=center,font=\scriptsize]

\lstset{language=boogie,numbers=none,basicstyle=\scriptsize\ttfamily}

  \node[state] (head) {%
\begin{lstlisting}
head: assert I;
      assume fI;
      assert J;
      goto body, end;
\end{lstlisting}
};

  \node[state,below=of head] (body) {%
\begin{lstlisting}
body: assume b;
      B
      goto head;
\end{lstlisting}
};

  \node[state,left=of head] (end) {%
\begin{lstlisting}
end: assume ! b;
     E
\end{lstlisting}
};

\draw ($(head.south)+(-4mm,0mm)$) to ($(body.north)+(-4mm,0)$);
\draw ($(body.north)+(4mm,0)$) to ($(head.south)+(4mm,0mm)$);
\draw (head) to (end);

\node (lb) at ($(body.south)+(17mm,-2mm)$) {};
\node[above=36mm of lb] (lt) {};
\draw [dotted,-] (lb) to (lt);

  \node[state,right=30mm of head] (head2) {%
\begin{lstlisting}
head: while (b)
        invariant I;
        free invariant fI;
        invariant J;
      { B }
      goto end;
\end{lstlisting}
};

  \node[state,left=of head2] (end2) {%
\begin{lstlisting}
end: E
\end{lstlisting}
};

\draw (head2) to (end2);
\end{tikzpicture}
\caption{Transformation of loops from unstructured (left) to structured (right).}
\label{fig:loop-cgf-trans}
\end{figure}
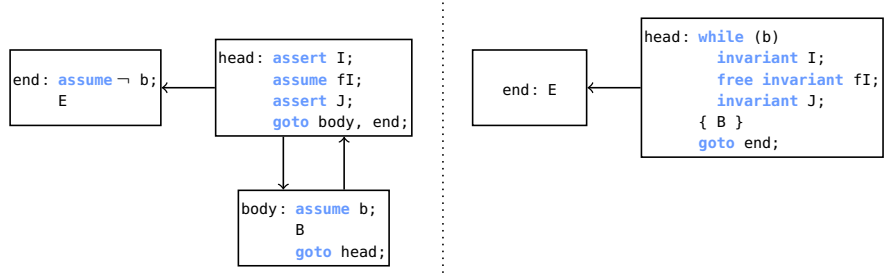

Consider the control-flow graph $G$ of a procedure body; each node $N$ of $G$ is a \emph{simple block}: a linear piece of code with a label $\ell_N$ on the first statement, no labels anywhere else in $N$, and a \B{goto} as last statement or no \B{goto} statements at all; arrows connect $N$ to the locations mentioned in $N$'s \B{goto} statement (if $N$ has no \B{goto}, we call it a \emph{terminal} node).
We apply three kinds of transformations on $G$ exhaustively.
\begin{description}
\item[Sequencing:] if $N \to M$ is the only arrow out of $N$ and the only arrow into $M$, and $M \not\to N$, replace $N$ and $M$ with the single block \B{$N$;$M$} with the \B{goto} at the end of $N$ and label $\ell_M$ removed.

\item[Choosing:] if $N \to \{M_1, \ldots, M_n\}$ are the only arrows out of $N$ and the only arrows into each $M_1, \ldots, M_n$, and every $M_k$, for $1 \leq k \leq n$, is a terminal node, replace $N, M_1, \ldots, M_n$ with the single block:
\begin{center}
\B{$N$; if (*) $\openCurl M_1 \closeCurl\:$ else $\:\openCurl$if (*) $\openCurl M_2 \closeCurl\:$ else $\:\openCurl\cdots\:$ else $\:\openCurl M_n \closeCurl\closeCurl \cdots \closeCurl$} 
\end{center}
with the \B{goto} at the end of $N$ and all labels other than $\ell_N$ removed.\footnote{This is after Dafny's calculational proof approach \cite{LeinoP13}.}

\item[Looping:] replace the subgraph of \autoref{fig:loop-cgf-trans} (left) with the structured loop to its right.
\end{description}
\fi

\iflong
\tightParagraph{Conditionals.}
The translation of conditionals is straightforward:
\begin{equation*}
\tr(\MB{if (b) then \{BT\} else \{BE\}}) \ =\ 
\MW{if}\ \tr(\MB{b})\ \MW{then \{}\ \tr(\MB{BT})\ \MW{\} else \{}\ \tr(\MB{BE})\;\MW{\}}
\end{equation*}
\fi

\tightParagraph{Loops.}
\autoref{fig:tr-loops-while} shows the translation of a \Boogie loop into a \WhyML loop.
An invariant marked as \B{free} can be assumed but need not be checked; correspondingly, the translation adds assumptions that ensure it holds at loop entrance and after every iteration.
The exception handling block surrounding the loop in \WhyML emulates the semantics of the control-flow breaking statement \B{break}: $\tr(\MB{break}) = \MW{raise Break}$.

\begin{figure}
\lstset{xleftmargin=4mm}
\centering
\footnotesize
\begin{tabular}{m{.4\textwidth} m{.4\textwidth}}
{\begin{lstlisting}[language=boogie]
while (b)
 invariant I;
 free invariant fI;
{ B }
\end{lstlisting}} &
{\begin{lstlisting}[language=Why3]
assume { $\tr($fI$)$ }
try while $\tr($b$)$ do
  invariant { $\tr($I$)$ }
  invariant { $\tr($fI$)$ }
  $\tr($B$)$
  assume { $\tr($fI$)$ }
done;
with | Break -> assume { $\tr($fI$)$ } end
\end{lstlisting}}
\end{tabular}
\caption{Translation of a \Boogie loop (left) into \WhyML (right).}
\label{fig:tr-loops-while}
\end{figure}

\iflong
\tightParagraph{Procedure calls.}
The translation of procedure calls is straightforward; for \Boogie procedure \B{p} in \autoref{fig:tr-procedures}:
$
\tr(\MB{call r := p(e)}) \ =\ 
\tr(\MW{t})\;\MW{ <- p(}\tr(\MW{e})\MW{)}
$.
Since \WhyML function calls  translating \Boogie procedures use the \W{val} style of declaration rather than the recursive function style (\W{rec}), the modular semantics of procedure calls (where the behavior is entirely determined by the specification) is correctly preserved.
\fi

\iflong
\tightParagraph{Call-forall.}
$\tr$ translates call-forall statements (supported in older versions of \Boogie~\cite{BoogieManual}) by axiomatizing their semantics:
\begin{center}
\B{$\des($\\cforall $\;$Lemma(*)$)\ =\ $}\B{ assume (forall t: T :: R$($t$)$ ==> E$($t$)$)}
\end{center}
where \B{Lemma} is declared as \B{procedure Lemma(t: T) requires R$($t$)$; ensures E$($t$)$}.
\fi

\iflong
\feature{Attributes}
$\tr$ translates \emph{triggers} using \WhyML's syntax: 
\begin{center}
\B{$\tr($forall x: X :: $\;\openCurl$trig$\closeCurl\;$ E(x)$) =\ $}\W{forall x: $\tr($X$)$ [$\tr($trig$)$]. $\tr($E(x)$)$}
\end{center}
The translation discards other application-specific attributes, which have no equivalent in \Why.
\fi

\iflong
\feature{Identifiers and Visibility}
\Boogie is more liberal than \WhyML in the range of characters that are allowed in identifier names; therefore, the translation defines an injective renaming of identifiers when necessary.

\Boogie allows local declarations to shadow global declarations of entities with the same name.
Since \WhyML does not allow shadowing, the translation introduces fresh names for local declarations when necessary to avoid name clashes with the shadowed declarations.

While the order of declarations is immaterial in \Boogie, in \WhyML reference must follow declaration.
Thus, the translation reorders declarations to comply with \WhyML's requirements; it also introduces a canonical order of declarations: types, global variables, functions, axioms, procedure declarations (\W{val}), procedure definitions (\W{let}), other declarations.

\fi

\feature{Polymorphic Maps} \label{sec:poly-maps-all}

We now consider \emph{polymorphic map} types, declared in \Boogie as:
\begin{equation}
\MB{type pM = <<}\vec{\alpha}\MB{>> [T}_1\MB{,} \ldots \MB{,T}_n\MB{] U}
\label{eq:def-poly-map}
\end{equation}
where $\vec{\alpha}$ is a vector $\alpha_1,\ldots,\alpha_m$ of $m > 0$ type parameters, and some of the types $\MB{T}_1, \ldots, \MB{T}_n, \MB{U}$ in \B{pM}'s definition depend on $\vec{\alpha}$. 
In the next paragraph, we explain why polymorphic maps cannot be translated to \WhyML directly.
Instead, we replace them with several monomorphic maps based on a global analysis of the types that are actually used in the \Boogie program being translated.
The result of this rewrite is a \Boogie program without polymorphic maps, which we can translate to \Why following the rules we previously described.
The shortcoming of this approach is that it gives up \emph{modularity}: verification holds only for the concrete types that are used (closed-word assumption); this seems to be necessary to express \Boogie's extremely liberal polymorphism without resorting to intricate ``semantic'' translations, which would likely fail verifiability.

\subsubsection{\Boogie vs.\ \WhyML polymorphism.}
While \WhyML also supports generic polymorphism\iflong, like every functional language in the ML family to which it belongs\fi{}, its usage is more restrictive than \Boogie's.
The first difference is that \emph{mutable} maps cannot be polymorphic in \WhyML\iflong; therefore, \Boogie variables of polymorphic map type require a special translation\fi.
The second difference is that, in some contexts, a variable of polymorphic map type in \Boogie effectively corresponds to \emph{multiple} maps\iflong, one for each possible concrete type, and the different maps can be combined in the same expression\fi.
Consider, for example, a 
\B{type Mix = <<\\alpha>>[\\alpha]\\alpha} of maps from generic\iflong{} type\fi{} $\alpha$ to $\alpha$; \Boogie accepts formulas such as \B{axiom (forall m: Mix :: m[0] == 1 && m[true])} where \B{m} acts as a map over \B{int} in the first conjunct and as a map over \B{bool} in the second\iflong conjunct\fi.
\WhyML, in contrast, always makes the type parameters explicit; hence, a logic variable of type \W{map 'a 'a} denotes a single map of a generic type that can only feature in expressions which do not assume anything about the concrete type that will instantiate~\W{'a}.
{}\iflong
Note that \Boogie even allows expressions that introduce inconsistencies, such as \B{forall <<\\beta>> x: \\beta, y: Mix :: y[x] == 3 && y[x] == true} (where the quantification is also type-generic), which passes typechecking but allows one to derive false.
\fi

Besides type declarations and quantifications, polymorphic maps can appear within polymorphic functions and procedures, declared as: 
\begin{gather}
\MB{function pF<<}\vec{\alpha}\MB{>>(x}_1\MB{: T}_1\MB{,} \ldots \MB{,x}_n\MB{: T}_n\MB{) returns (U)}
\label{eq:def-poly-fun} \\
\MB{procedure pP<<}\vec{\alpha}\MB{>>(x}_1\MB{: T}_1\MB{,} \ldots \MB{,x}_n\MB{: T}_n\MB{) returns (u: U)}
\label{eq:def-poly-proc}
\end{gather}
\iflong
Precisely, two kinds of polymorphic maps may feature within polymorphic functions and procedures: polymorphic maps generic with respect to \emph{explicitly declared} function or procedure parameters are similar to \Why's, and hence different from those generic with respect to implicit type parameters declared outside the function or procedure.
For example, implementations of a procedure \B{p<<\\beta>>(m: Mix, n: [\\beta]\\beta)} can select elements of any concrete type from \B{m}, but only elements of parametric type $\beta$ from \B{n}.
\fi

\subsubsection{Type analysis.}
We have seen that a \Boogie polymorphic map may correspond to multiple monomorphic maps in certain contexts.
The translation reifies this idea based on global type analysis: for every item (constant, program or logic variable, or formal argument) \B{pm} of polymorphic map type \B{pM} as in \eqref{eq:def-poly-map}, it determines the set $\typesOf{pm}$ of all actual types \B{pm} takes in expressions or assignments, as outlined in \autoref{tab:type-analysis}.\iflong\footnote{A parameter's actual type is ambiguous if the parameter appears in the map type's codomain but not in its domain; in this case, \Boogie defaults to type \B{int}.}\fi{} 
This in turn determines the set $\typesOf{pM}$ as the union of all sets $\typesOf{p}$ for \B{p} of type \B{pM}.

\begin{table}[!hbt]
\setlength{\tabcolsep}{2pt}
\centering
\scriptsize
\begin{tabular}{clll}
& &  &  $\typesOf{pm}$ includes $\MB{[}t_1\MB{,} \ldots\MB{,} t_n\MB{]}u$ such that: \\
\hline
\multirow{4}{*}{expressions} 
& read   &  $\MB{pm}$  & $\MB{pm} :: \MB{[}t_1\MB{,} \ldots\MB{,} t_n\MB{]}u$ \\
& select &  $\MB{pm[e}_1\MB{,}\ldots\MB{,e}_n\MB{]}$ & $\MB{e}_1 :: t_1, \ldots, \MB{e}_n :: t_n, \MB{pm[e}_1\MB{,}\ldots\MB{,e}_n\MB{]} :: u$ \\
& update & $\MB{pm[e}_1\MB{,}\ldots\MB{,e}_n\MB{:= f]}$ & $\MB{e}_1 :: t_1, \ldots, \MB{e}_n :: t_n, \MB{f} :: u$ \\
& function reference &  $\MB{f(it)}$ & $\MB{it} :: \MB{[}t_1\MB{,} \ldots\MB{,} t_n\MB{]}u$, where $\MB{function f(pm: pM)}$ \\
\cline{2-4}
\multirow{5}{*}{statements} 
& copy & $\MB{pm} := \MB{it}$ &  $\MB{it} :: \MB{[}t_1\MB{,} \ldots\MB{,} t_n\MB{]}u$  \\
& assignment & $\MB{pm[e}_1\MB{,}\ldots\MB{,e}_n\MB{]} := \MB{f}$ & $\MB{e}_1 :: t_1, \ldots, \MB{e}_n :: t_n, \MB{f} :: u$ \\
& havoc & $\MB{havoc pm}$ & -- \\
& procedure call in &  $\MB{call}\;\MB{p(it)}$ & $\MB{it} :: \MB{[}t_1\MB{,} \ldots\MB{,} t_n\MB{]}u$, where $\MB{procedure p(pm: pM)}$ \\
& procedure call out &  $\MB{call it :=}\;\MB{p()}$ & $\MB{it} :: \MB{[}t_1\MB{,} \ldots\MB{,} t_n\MB{]}u$, where $\MB{procedure p() returns(pm: pM)}$
\end{tabular}
\caption{Each occurrence of an item \B{pm} of polymorphic map type \B{pM} determines the set $\typesOf{pm}$ of actual types. ($\MB{x} :: t$ denotes that \B{x} has type $t$.)}
\label{tab:type-analysis}
\end{table}

The types in $\typesOf{pM}$ include in general both concrete and parametric types.
For example, the program of \autoref{fig:ex-types-analysis} (left) determines $\typesOf{m} = \{ \MB{[int]int}, \MB{[\\beta]\\beta} \}$, \linebreak $\typesOf{n} = \{ \MB{[bool]bool} \}$, and $\typesOf{M} = \typesOf{m} \cup \typesOf{n}$, 
where $\beta$ is procedure \B{p}'s type parameter (since \B{p} is not called anywhere, that's the only known actual type of \B{x}).
Let $\ctypesOf{pM}$ denote the set of all \emph{concrete} types in $\typesOf{pM}$.

\begin{figure}[!htb]
\lstset{xleftmargin=0mm}
\scriptsize
\begin{tabular}{p{.35\textwidth} p{.5\textwidth}}
{\begin{lstlisting}[language=boogie]
type M = <<\alpha>> [\alpha]\alpha;
var m: M;
axiom (forall n: M :: n[true]);


procedure p<<\beta>>(x: \beta)
  requires (forall i: int :: m[i] == i);
  modifies m;
{ m[x] := x; }\end{lstlisting}}
&
{\begin{lstlisting}[language=boogie]
type (M_int, M_bool, M_a) = ([int]int, [bool]bool, [a]a);
var (m_int, m_bool, m_a): (M_int, M_bool, M_a);
axiom (forall (n_int, n_bool, n_a): (M_int, M_bool, M_a) ::
         n_bool[true]);

procedure (p_int, p_bool, p_a)(x: (int, bool, a))
  requires (forall i: int :: m_int[i] == i);
  modifies (m_int, m_bool, m_a);
{ (m_int, m_bool, m_a)[x] := x }\end{lstlisting}}
\end{tabular}
\caption{An example of how polymorphic maps (left) translate to monomorphic (right).
Procedure~\B{p} translates to 3 procedures \W{p_int}, \W{p_bool}, and \W{p_a}, each with argument of type \W{int}, \W{bool}, or \W{a}.}
\label{fig:ex-types-analysis}
\end{figure}

\subsubsection{Desugaring polymorphic maps.}
To describe how the translation replaces polymorphic maps by monomorphic maps, we introduce a pseudo-code notation that allows \emph{tuples} (in round brackets) of program elements where normally only a single element is allowed.
The semantics of this notation corresponds quite intuitively to multiple statements or declarations.
For example, a variable declaration \B{var (x, y): (int, bool)} is a shorthand for declaring variables \B{x: int} and \B{y: bool}; a formula \B{(x, y) == (3, true)} is a shorthand for \B{x == 3 && y}; and a procedure declaration using the tuple notation \B{procedure (p_int, p_bool)(x: (int, bool))} is a shorthand for declaring two procedures \B{p_int(x: int)} and \B{p_bool(x: bool)}.

We also use the following notation: given an $n$-vector $\vec{a} = a_1, \ldots, a_n$ and a type expression $T$ parametric with respect to $\vec{\alpha}$, $T_{\vec{a}}$ denotes $T$ with $a_k$ substituted for $\alpha_k$, for $k = 1, \ldots, n$.
If $\BBT$ is a set of types obtained from the same type expression $T$, such as $\typesOf{pM}$ with respect to \B{pM}'s definition, and \B{id} is an identifier, let $\MB{(}\BBT\MB{)}$ denote $\BBT$ as a tuple, and $\MB{(id_}\BBT\MB{)}$ denote the tuple of identifiers $\MB{id_}\vec{t}$ such that $T_{\vec{t}}$ is the corresponding type in $\BBT$.
In the example of \autoref{fig:ex-types-analysis}, if $T = \MB{[\\alpha]\\alpha}$ then $T_{\MB{int}} = \MB{[int]int}$, $\MB{(}\typesOf{m}\MB{)} = (\MB{[int]int,} \MB{[\\beta]\\beta})$, and $\MB{(j_}\typesOf{m}\MB{)} = \MB{(j_int, j_\\beta)}$.
Throughout, we also assume that an uninterpreted type \B{a$_k$} is available for $k = 1, \ldots, n$, that $M_{\MB{a}}$ denotes the type expression $\MB{[T}_1\MB{,} \ldots \MB{,T}_n\MB{] U}$ in \eqref{eq:def-poly-map} with each $\alpha_k$ replaced by \B{a$_k$}, and that $\concPlus{pM} = \ctypesOf{pM} \cup \{M_{\MB{a}}\}$.

\tightParagraph{Declarations.}
Type declaration \eqref{eq:def-poly-map} desugars to several type declarations:
\begin{equation}
\MB{type (pM_}\concPlus{pM}\MB{)) = (}\concPlus{pM}\MB{)}
\label{eq:poly-map-desugared}
\end{equation}
The declaration of an \emph{item} \B{pm: pM}, where \B{pm} can be a constant, or a program or logic variable,  desugars to a declaration  
$\MB{(pm_}\concPlus{pM}\MB{)): (}\concPlus{pM}\MB{)}$ of multiple items of the same kind.
The declaration of a \emph{procedure} or \emph{function} \B{g} with an (input or output) argument \B{x: pM} desugars to a declaration of multiple procedures or functions 
$\MB{(g_}\concPlus{pM}\MB{)}\MB{(x: (}\concPlus{pM}\MB{)}$---multiple declarations each with one variant of \B{x}
; if \B{g} has multiple arguments of this kind, the desugaring is applied recursively to each variant.
\autoref{fig:ex-types-analysis} (right) shows how the polymorphic map type \B{M} and each of the items \B{m} and \B{n} of type \B{M} become 3 monomorphic types and 3 items of these monomorphic types.

For every polymorphic function or procedure \B{g} with type parameters $\vec{\beta}$, also consider any one of their arguments declared as $\MB{x}\colon X$.
If $X$ is a type expression that depends on $\vec{\beta}$, and there exists a map type $\MB{[}V_1\MB{,}\ldots\MB{,}V_n\MB{]}V_0$ in $\typesOf{pM}$ such that $X = V_k$ for some $k = 0, \ldots, n$, then \B{g} becomes $\MB{(g_}\mathbb{V}_k\MB{)(x: (}\mathbb{V}_k\MB{))}$---corresponding to multiple \B{g}'s each with one argument, 
where $\mathbb{V}_k = \left\{ \overline{V}_k \mid \MB{[}\overline{V}_1\MB{,}\ldots\MB{,}\overline{V}_n\MB{]}\overline{V}_0 \in \concPlus{pM} \right\}$ is the set of all concrete types that instantiate the $k$th type component.
This transformation enables assigning arguments to polymorphic maps inside polymorphic functions or procedures that have become monomorphic.
\autoref{fig:ex-types-analysis} (right) shows how argument \B{x: \\beta} becomes an argument of concrete type \B{int}, \B{bool}, or \B{a}, since $\MB{[\\beta]\\beta} \in \typesOf{M}$.
(As procedure \B{p} does not use $\beta$ elsewhere, we drop it from the signature.)

\tightParagraph{Expressions.}
Every occurrence---in expressions, as l-values of assignments, and as targets of \B{havoc} statements---of an item \B{w} of polymorphic type \B{W} whose declaration has been modified to remove polymorphic map types is replaced by one or more of the newly introduced monomorphic types as follows.
If \B{w}'s actual type within its context is a \emph{concrete} type \B{C}, then we replace \B{w} with $\MB{w_}c$ such that $\MB{W}_c = \MB{C}$; otherwise, \B{w}'s actual type is a \emph{parametric} type, and we replace \B{w} with the tuple $\MB{(w_}X\MB{)}$, including all variants of \B{w} that have been introduced.
In \autoref{fig:ex-types-analysis} (right), \B{n[true]} rewrites to just \B{n_bool[true]} since the concrete type is \B{bool}; the assignment in \B{p}'s body, whose actual type is parametric with respect to $\beta$, becomes an assignment involving each of the three variants of \B{m} corresponding to the three variants of \B{p} that have been introduced.

\section{Implementation and Experiments}\label{sec:experiments}

\subsection{Implementation}
We implemented the translation $\tr$ described in \autoref{sec:translation} as a command-line tool \btw implemented in Java~8.
\btw works as a staged filter: 1) it parses and typechecks the input \Boogie program, and creates a \Boogie AST (abstract syntax tree); 2) it desugars the \Boogie AST according to $\des$; 3) it transforms the \Boogie AST into a \WhyML AST according to $\enc$; 4) it outputs the \WhyML AST in the form of code. 

Stage 1) relies on Sch\"af's parsing and typechecking library Boogieamp\footnote{\url{https://github.com/martinschaef/boogieamp}}, which we modified to support access using the visitor pattern, AST in-place modifications, and the latest syntax of \Boogie (e.g., for integer vs.\ real division\iflong\footnote{\url{http://boogie.codeplex.com/discussions/397357}}\fi).
Stages 2) and 3) are implemented by multiple AST visitors, each taking care of a particular aspect of the translation, in the style of~\cite{TFNMO12-WCRE12}; the overhead of traversing the AST multiple times is negligible and improves modularity: handling a new construct (for example, in future versions of \Boogie) or changing the translation of one feature only requires adding or modifying one feature-specific visitor class.{} \iflong{A similar technique is also advocated in~\cite{SarkarWD05}.}\fi

\subsection{Experiments}

The goal of the experiments is ascertaining that \btw can translate realistic \Boogie programs producing \WhyML programs that can be verified taking advantage of \Why's multiple back-end support.
The experiments are limited to fully-automated verification, and hence do not evaluate other possible practical benefits of translating programs to \WhyML such as support for interactive provers and executability for testing purposes.

\tightParagraph{Programs.}
The experiments target a total of 194 \Boogie programs from three groups according to their origin: group \nat (native) includes 29 programs that encode algorithmic verification problems directly in \Boogie (as opposed to translating from a higher-level language); group \obj (object-oriented) includes 6 programs that are based on a heap-based memory model; group \tes (tests) includes 159 programs from \Boogie's test suite. \autoref{tab:programs-loc} summarizes the sizes of the programs in each group.

\begin{table}
\centering
\scriptsize
\setlength{\tabcolsep}{3pt}
\begin{tabular}{l r |*{1}{rrrr} |*{1}{rrrr}}
\multicolumn{1}{c}{} &
\multicolumn{1}{c}{} &
\multicolumn{4}{c}{\textsc{loc \Boogie}} &
\multicolumn{4}{c}{\textsc{loc \WhyML}}
\\
\multicolumn{1}{c}{\textsc{group}} &
\multicolumn{1}{c}{\textsc{\#}} &
\multicolumn{1}{c}{$m$} &
\multicolumn{1}{c}{$\mu$} &
\multicolumn{1}{c}{$M$} &
\multicolumn{1}{c}{$\Sigma$} &
\multicolumn{1}{c}{$m$} &
\multicolumn{1}{c}{$\mu$} &
\multicolumn{1}{c}{$M$} &
\multicolumn{1}{c}{$\Sigma$} 
\\
\hline
\nat  &  29  &  20  &  73  &  253  &  2110  &  62  &  128  &  318  &  3716
 \\
\obj  &  6  &  44  &  146  &  385  &  878  &  90  &  208  &  446  &  1245
 \\
\tes  &  159  &  3  &  21  &  155  &  3272  &  36  &  64  &  290  &  10180
 \\
\hline
\textbf{Total}:  &  194  &  3  &  34  &  385  &  6260  &  36  &  106  &  446  &  15141
 \\

\end{tabular}
\caption{A summary of the \Boogie programs used in the experiments, and their translation to \WhyML using \btw. For each program \textsc{group}, the table reports how many programs it includes (\textsc{\#}), the minimum $m$, mean $\mu$, maximum $M$, and total $\Sigma$ length in non-comment non-blank lines of code (\textsc{loc}) of those \textsc{\Boogie} programs and of their \textsc{\WhyML} translations.}
\label{tab:programs-loc}
\end{table}

The programs in \nat, which we developed in previous work \cite{FMV-CSUR14,F14-TR-20062014}, include several standard algorithms such as sorting and array rotation.
The programs in \obj include 2 simple examples in Java and 1 in Eiffel, encoded in \Boogie by Joogie~\cite{ArltS12} and AutoProof~\cite{TFNP-TACAS15} (we manually simplified AutoProof's translation to avoid features \btw doesn't support), and 3 algorithmic examples adapted from \nat to use a global heap in the style of object-oriented programs.
Among the 515 programs that make up \Boogie's test suite\footnote{\url{https://github.com/boogie-org/boogie/tree/master/Test}} we retained in \tes those that mainly exercise features supported by \btw.
\iflong
This meant excluding several groups of tests that exercise special options (Houdini, assertion inference, special Z3 encodings and directives, etc.), unsupported language features (bitvectors, gotos, etc.), and the correctness of typechecking (\btw assumes well-formed \Boogie input).
It also meant excluding 4 programs that triggered \Boogie errors (a \Boogie \emph{error} means here a problem with the input such as a typechecking or parsing error due to a feature not activated; it is not a verification error, which just denotes a failed verification attempt and is fair game for evaluating the translation); and another 35 programs that \btw failed to translate because of unsupported features that we identified a posteriori.
\fi

\tightParagraph{Setup.}
Each experiment targets one \Boogie program \verb|b|: it runs \Boogie with command \verb|boogie b| and a timeout of 180 seconds; it runs \btw to translate \verb|b| to \verb|w| in \WhyML; for each SMT solver \verb|p| among Alt-Ergo, CVC3, CVC4, and Z3, it runs \Why with command \verb|why3 prove -P p w|, also with a timeout of 180 seconds.\iflong\footnote{The timeouts were enforced using the Unix command \texttt{timeout}. We also set had a 20-second timeout per procedure (option \texttt{/timeLimit} in \Boogie) or goal (option \texttt{-T} in \Why).}\fi{} 
For each run we collected the wall-clock running time, the total number of verification goals, and how many of such goals the tool verified successfully.\iflong\footnote{The number of verification goals of each program is the same in \Boogie and \Why: the number of procedure implementations.}\fi 

All the experiments ran on a Ubuntu 14.04 LTS GNU/Linux box with 8-core Intel i7-4790 CPU at 3.6~GHz and 16~GB of RAM, with the following tools: Alt-Ergo~0.99.1, CVC3~2.4.1, CVC4~1.4, Z3~4.3.2, Mono~4.2.2, \iflong OCaml~4.02.3,\fi{} Boogie~2.3.0.61016, and Why3~0.86.2.
To account for noise, we repeated each verification three times and report 
the mean value of the 95th percentile 
of the running times.

\begin{table}
\centering
\scriptsize
\setlength{\tabcolsep}{4pt}
\begin{tabular}{l r |*{3}{r}|*{3}{r}|*{1}{r}}
\multicolumn{1}{c}{\textsc{group}} &
\multicolumn{1}{c}{\textsc{\#}} &
\multicolumn{1}{c}{\textsc{b = w}} &
\multicolumn{1}{c}{\textsc{b > w}} &
\multicolumn{1}{c}{\textsc{b < w}} &
\multicolumn{1}{c}{\textsc{0=0}} &
\multicolumn{1}{c}{\textsc{50=50}} &
\multicolumn{1}{c}{\textsc{100=100}} &
\multicolumn{1}{c}{\textsc{spurious}} 
\\
\hline
\nat  &  29  &  19  &  10  &  0  &  1  &  0  &  18  &  0
 \\
\obj  &  6  &  5  &  0  &  1  &  1  &  2  &  2  &  0
 \\
\tes  &  159  &  137  &  21  &  1  &  71  &  21  &  45  &  0
 \\
\hline
\textbf{Total}:  &  194  &  161  &  31  &  2  &  73  &  23  &  65  &  0
 \\

\end{tabular}
\caption{A summary of how \Boogie performs in comparison with \Why. For each program \textsc{group}, the table reports
how many programs it includes (\textsc{\#}), 
for how many of the programs \Boogie verifies as many goals (\textsc{b = w}), more goals (\textsc{b > w}), or fewer goals (\textsc{b < w}) than \Why with any of the SMT solvers; 
for how many of the programs both \Boogie and \Why verify none (\textsc{0=0}), some but not all (\textsc{50=50}), or all (\textsc{100=100}) of the goals; 
the last column (\textsc{spurious}) indicates that \btw's translation never introduces spurious goals that are proved by \Why (that is, if \Boogie's input has zero goals, so does \WhyML's translation).}
\label{tab:programs-val}
\end{table}

\tightParagraph{Results.}
\autoref{tab:programs-val} shows a summary of the results where we compare \Why's best performance, with any one of the four SMT solvers, against \Boogie's.
The most significant result is that the \WhyML translation produced by \btw behaves like the \Boogie original in 83\% (161, \textsc{b=w}) of the experiments.
This means that \Boogie may fail to verify all goals (column \textsc{0=0}), verify some goals and fail on others (column \textsc{50=50}), or verify all goals (column \textsc{100=100}); in each case, \Why consistently verifies the same goals on \btw's translation.
Indeed, many programs in \tes are tests that are supposed to fail verification; hence, the correct behavior of the translation is to fail as well.
We also checked the failures of programs in \nat and \obj to ascertain that \btw's translation preserves correctness.
\autoref{tab:programs-val} does not show this, but we also found another 2 programs in \nat \iflong(\verb|inv_survey/bst| and \verb|rotation/rotation_reverse|)\fi{} where \Why proves the same goals as \Boogie only by combining the results of multiple SMT solvers.

\Boogie verifies more goals than \Why in 16\% (31, \textsc{b > w}) of the experiments, where it is more effective because of better features (default triggers, invariant inference, SMT encoding) or simply because of some language features that are not fully supported by \btw (examples are Z3-specific annotations, which \btw simply drops, and \B{goto}, which \btw encodes as \B{assert false} to ensure soundness).
In 1\% (2, \textsc{b < w}) of the experiments, \Why even verifies more goals than \Boogie.
One program in \obj \iflong(\verb|rotation_by_copy|)\fi{} is a genuine example where \Why's Z3 encoding is more effective than \Boogie's\iflong\footnote{However, \Boogie also succeeds given a longer timeout thank the one used in the experiments.}\fi{}; the one program in \tes \iflong(\verb|test2/Quantifiers|)\fi{} should instead be considered spurious, as it deploys some trigger specifications that are \Boogie-specific (negated triggers) or interact in a different way with the default triggers.
\iflong
(Procedures \B{S}, \B{U0}, and \B{U1} use regular triggers whose translation to \Why yields a different behavior, probably because of \Why's default triggers differ from \Boogie's; procedures \B{W} and \B{X2} use negated triggers, which \btw ignore.)
\fi
As this was the only program in our experiments that introduced clearly spurious behavior, the experiments provide convincing evidence that \btw's translation preserves correctness and verifiability to a large degree.

\begin{table}
\centering
\scriptsize
\begin{adjustwidth}{-8mm}{-8mm}
\begin{tabular}{l r *{1}{|rrr|rrr} *{4}{||rrr|rrr}}
\multicolumn{1}{c}{} &
\multicolumn{1}{c}{} &
\multicolumn{6}{c}{\textsc{Z3 \Boogie}} &
\multicolumn{6}{c}{\textsc{Alt-Ergo \Why}} &
\multicolumn{6}{c}{\textsc{CVC3 \Why}} &
\multicolumn{6}{c}{\textsc{CVC4 \Why}} &
\multicolumn{6}{c}{\textsc{Z3 \Why}}
\\
\multicolumn{1}{c}{} &
\multicolumn{1}{c}{} &
\multicolumn{3}{c}{\textsc{outcome}} &
\multicolumn{3}{c}{\textsc{time}} &
\multicolumn{3}{c}{\textsc{outcome}} &
\multicolumn{3}{c}{\textsc{time}} &
\multicolumn{3}{c}{\textsc{outcome}} &
\multicolumn{3}{c}{\textsc{time}} &
\multicolumn{3}{c}{\textsc{outcome}} &
\multicolumn{3}{c}{\textsc{time}} &
\multicolumn{3}{c}{\textsc{outcome}} &
\multicolumn{3}{c}{\textsc{time}}
\\
\multicolumn{1}{c}{\textsc{group}} &
\multicolumn{1}{c}{\textsc{\#}} &
\multicolumn{1}{c}{$\mu$} &
\multicolumn{1}{c}{$\forall$} &
\multicolumn{1}{c}{$\not\exists$} &
\multicolumn{1}{c}{$\mu$} &
\multicolumn{1}{c}{$\Sigma$} &
\multicolumn{1}{c}{$\infty$} &
\multicolumn{1}{c}{$\mu$} &
\multicolumn{1}{c}{$\forall$} &
\multicolumn{1}{c}{$\not\exists$} &
\multicolumn{1}{c}{$\mu$} &
\multicolumn{1}{c}{$\Sigma$} &
\multicolumn{1}{c}{$\infty$} &
\multicolumn{1}{c}{$\mu$} &
\multicolumn{1}{c}{$\forall$} &
\multicolumn{1}{c}{$\not\exists$} &
\multicolumn{1}{c}{$\mu$} &
\multicolumn{1}{c}{$\Sigma$} &
\multicolumn{1}{c}{$\infty$} &
\multicolumn{1}{c}{$\mu$} &
\multicolumn{1}{c}{$\forall$} &
\multicolumn{1}{c}{$\not\exists$} &
\multicolumn{1}{c}{$\mu$} &
\multicolumn{1}{c}{$\Sigma$} &
\multicolumn{1}{c}{$\infty$} &
\multicolumn{1}{c}{$\mu$} &
\multicolumn{1}{c}{$\forall$} &
\multicolumn{1}{c}{$\not\exists$} &
\multicolumn{1}{c}{$\mu$} &
\multicolumn{1}{c}{$\Sigma$} &
\multicolumn{1}{c}{$\infty$}
\\
\hline
\nat  &  29  &  93  &  25  &  1  &  0.4  &  12  &  0  &  61  &  14  &  6  &  20.6  &  598  &  0  &  28  &  1  &  12  &  0.2  &  5  &  0  &  33  &  2  &  11  &  30.1  &  873  &  0  &  73  &  16  &  5  &  12.6  &  367  &  0
 \\
\obj  &  6  &  52  &  2  &  2  &  3.9  &  23  &  0  &  46  &  1  &  2  &  30.1  &  181  &  0  &  46  &  1  &  2  &  0.2  &  1  &  0  &  52  &  2  &  2  &  28.4  &  170  &  0  &  68  &  3  &  1  &  23.7  &  142  &  0
 \\
\tes  &  159  &  45  &  55  &  71  &  0.3  &  53  &  0  &  37  &  45  &  85  &  25.8  &  4096  &  1  &  33  &  39  &  91  &  0.1  &  18  &  0  &  37  &  45  &  86  &  27.4  &  4360  &  1  &  37  &  44  &  86  &  25.9  &  4121  &  1
 \\
\hline
\textbf{Total}:  &  194  &  60  &  82  &  74  &  0.7  &  88  &  0  &  53  &  60  &  93  &  22.6  &  4875  &  1  &  30  &  41  &  105  &  0.2  &  24  &  0  &  35  &  49  &  99  &  29.7  &  5403  &  1  &  69  &  63  &  92  &  14.5  &  4630  &  1
 \\

\end{tabular}
\end{adjustwidth}
\caption{For each program \textsc{group} the table reports how many programs it includes (\textsc{\#}) and, for both \Boogie and \Why for each choice of SMT solver among \textsc{Alt-Ergo}, \textsc{CVC3}, and \textsc{Z3}: the mean percentage of goals verified in each program (\textsc{outcome} $\mu$), how many programs were completely verified (\textsc{outcome} $\forall$), and how many were not verified at all (\textsc{outcome} $\not\exists$), the mean $\mu$ and total $\Sigma$ verification \textsc{time} in seconds\iflong{} (including time outs)\fi, and how many programs timed out.}
\label{tab:programs-ver}
\end{table}

\autoref{tab:programs-ver} provides data about the experiments' running times, and differentiates the performance of each SMT solver with \Why.
Z3 is the most effective SMT solver in terms of programs it could completely verify (columns $\forall$), followed by Alt-Ergo.
While CVC3 is generally the least effective, it has the advantage of returning very quickly (only 0.2 seconds of average running time), even more quickly than Z3 in \Boogie. 
CVC4 falls somewhere in the middle, in terms both of effectiveness and of running time.
\Boogie's responsiveness remains excellent if balanced against its effectiveness; a better time-effectiveness of \Why with Alt-Ergo and Z3 could be achieved by setting tight per-goal timeouts (in most cases, verification attempts that last longer than a few seconds do not eventually succeed).

\section{Discussion}\label{sec:conclusions}

The current implementation of the translation $\tr$ has some limitations that somewhat restrict its applicability.
As we already mentioned in the paper, some features of the \Boogie language are not supported (bitvectors, gotos), or only partially supported (polymorphic mappings); and frame specifications are assumed.
All of these are, however, limitations of the current prototype implementation only, and we see no fundamental hurdles to extending \btw along the lines of the definition of $\tr$ in \autoref{sec:translation}.

Since \btw also takes great care to confine the effect of translating \Boogie programs that include unsupported features, and to fail when it cannot produce a correct translation, it still largely preserves \emph{correctness} (soundness, in particular).
\iflong
For example, a \B{goto} statement is rendered as \B{assert false}; therefore, the translated program verifies only if the \B{goto} is never executed in the original program, which ensures soundness.
\fi{}
On the other hand, our experiments also demonstrate that the translation $\tr$, as implemented by \btw, largely meets the other goal of preserving \emph{verifiability}: even if the experimental subjects all are idiomatic \Boogie programs written independent of the translation effort, 83\% of the translated programs behave in \Why as they do in \Boogie.

In future work, we will address the features of \Boogie that are still not satisfactorily supported by \btw.
We will also devise strategies to take advantage of \Why's multi-prover support.
Other possible directions include formalizing the translation to prove that it preserves correctness; and devising a reverse translation from \WhyML to \Boogie.



\iflong

\newpage
\clearpage
\appendix

\begin{table}
\centering
\scriptsize
\begin{tabular}{l |r *{1}{rr} |r *{4}{rr}}
\multicolumn{1}{c}{} &
\multicolumn{3}{c}{\textsc{\Boogie}} &
\multicolumn{9}{c}{\textsc{\Why}}
\\
\multicolumn{1}{c}{} &
\multicolumn{1}{c}{} &
\multicolumn{2}{c}{\textsc{Z3}} &
\multicolumn{1}{c}{} &
\multicolumn{2}{c}{\textsc{Alt-Ergo}} &
\multicolumn{2}{c}{\textsc{CVC3}} &
\multicolumn{2}{c}{\textsc{CVC4}} &
\multicolumn{2}{c}{\textsc{Z3}}
\\
\multicolumn{1}{c}{\textsc{name}} &
\multicolumn{1}{c}{\textsc{loc}} &
\multicolumn{1}{c}{\textsc{\% v.}} &
\multicolumn{1}{c}{\textsc{t}} &
\multicolumn{1}{c}{\textsc{loc}} &
\multicolumn{1}{c}{\textsc{\% v.}} &
\multicolumn{1}{c}{\textsc{t}} &
\multicolumn{1}{c}{\textsc{\% v.}} &
\multicolumn{1}{c}{\textsc{t}} &
\multicolumn{1}{c}{\textsc{\% v.}} &
\multicolumn{1}{c}{\textsc{t}} &
\multicolumn{1}{c}{\textsc{\% v.}} &
\multicolumn{1}{c}{\textsc{t}}
\\
\hline
\verb|inv_survey/array_partitioning_v1|  &  42  &  100  &  0.5  &  100  &  50  &  20.1  &  50  &  0.2  &  50  &  21.3  &  50  &  20.3  \\
\verb|inv_survey/array_partitioning_v2|  &  53  &  100  &  0.4  &  125  &  100  &  0.3  &  50  &  0.1  &  50  &  21.3  &  100  &  0.1  \\
\verb|inv_survey/array_stack_reversal|  &  125  &  100  &  0.5  &  204  &  100  &  0.3  &  86  &  0.2  &  71  &  42.5  &  86  &  20.3  \\
\verb|inv_survey/bst|  &  153  &  100  &  0.4  &  258  &  50  &  40.5  &  50  &  0.2  &  75  &  21.4  &  75  &  20.3  \\
\verb|inv_survey/bubble_sort_basic|  &  49  &  100  &  0.5  &  113  &  100  &  0.2  &  50  &  0.1  &  50  &  21.3  &  100  &  0.2  \\
\verb|inv_survey/bubble_sort_improved|  &  53  &  100  &  0.5  &  118  &  100  &  0.6  &  50  &  0.2  &  50  &  21.3  &  100  &  0.1  \\
\verb|inv_survey/comb_sort|  &  56  &  100  &  0.4  &  124  &  100  &  0.3  &  50  &  0.2  &  50  &  21.3  &  100  &  0.2  \\
\verb|inv_survey/dutch_flag|  &  63  &  100  &  0.5  &  133  &  50  &  20.1  &  50  &  0.2  &  50  &  21.3  &  100  &  0.1  \\
\verb|inv_survey/insertion_sort|  &  47  &  100  &  0.6  &  100  &  0  &  20.1  &  0  &  0.1  &  0  &  21.3  &  100  &  0.2  \\
\verb|inv_survey/knapsack|  &  50  &  100  &  0.3  &  97  &  100  &  7.5  &  0  &  0.1  &  0  &  21.2  &  100  &  0.1  \\
\verb|inv_survey/Levenshtein_distance|  &  43  &  100  &  0.3  &  91  &  100  &  0.4  &  0  &  0.1  &  100  &  0.9  &  100  &  0.1  \\
\verb|inv_survey/max_of_array_v1|  &  20  &  100  &  0.4  &  66  &  100  &  0.1  &  0  &  0.1  &  0  &  21.2  &  100  &  0.1  \\
\verb|inv_survey/max_of_array_v2|  &  20  &  100  &  0.4  &  66  &  100  &  0.2  &  0  &  0.1  &  0  &  21.2  &  100  &  0.1  \\
\verb|inv_survey/partition|  &  63  &  100  &  0.4  &  137  &  100  &  0.8  &  50  &  0.1  &  50  &  21.3  &  100  &  0.1  \\
\verb|inv_survey/plateau|  &  43  &  100  &  0.5  &  84  &  0  &  20.1  &  0  &  0.1  &  0  &  21.2  &  0  &  20.3  \\
\verb|inv_survey/reverse|  &  68  &  100  &  0.4  &  131  &  100  &  0.3  &  0  &  0.1  &  0  &  21.3  &  100  &  0.1  \\
\verb|inv_survey/selection_sort|  &  72  &  100  &  0.4  &  160  &  100  &  4.9  &  33  &  0.2  &  33  &  42.4  &  100  &  0.2  \\
\verb|inv_survey/sequential_search_v1|  &  28  &  100  &  0.4  &  72  &  0  &  20.1  &  0  &  0.1  &  0  &  21.2  &  0  &  20.2  \\
\verb|inv_survey/sequential_search_v2|  &  23  &  100  &  0.4  &  70  &  0  &  20.1  &  0  &  0.1  &  0  &  21.2  &  0  &  20.2  \\
\verb|inv_survey/sum_of_array|  &  21  &  100  &  0.3  &  62  &  100  &  0.1  &  100  &  0.1  &  0  &  21.2  &  100  &  0.1  \\
\verb|inv_survey/welfare_crook|  &  44  &  100  &  0.3  &  86  &  100  &  0.1  &  0  &  0.1  &  100  &  0.8  &  0  &  20.2  \\
\verb|rotation/rotation_copy|  &  57  &  100  &  0.4  &  128  &  33  &  40.1  &  33  &  0.2  &  33  &  42.4  &  67  &  20.3  \\
\verb|rotation/rotation_copy_plain|  &  41  &  100  &  0.3  &  80  &  0  &  20.1  &  0  &  0.1  &  0  &  21.2  &  100  &  0.2  \\
\verb|rotation/rotation_reverse|  &  201  &  90  &  0.4  &  318  &  40  &  120.5  &  10  &  0.4  &  50  &  106.6  &  80  &  40.6  \\
\verb|rotation/rotation_swap-1_3|  &  48  &  0  &  0.3  &  88  &  0  &  20.1  &  0  &  0.1  &  0  &  21.2  &  0  &  20.2  \\
\verb|rotation/rotation_swap-2_3|  &  175  &  60  &  0.4  &  201  &  20  &  80.2  &  20  &  0.2  &  20  &  84.7  &  40  &  60.6  \\
\verb|rotation/rotation_swap-3_3|  &  47  &  100  &  0.3  &  96  &  67  &  20.1  &  67  &  0.2  &  67  &  21.3  &  100  &  0.2  \\
\verb|rotation/rotation_swap_iterative-1_2|  &  152  &  100  &  0.5  &  184  &  33  &  40.2  &  33  &  0.2  &  33  &  42.4  &  67  &  20.3  \\
\verb|rotation/rotation_swap_iterative-2_2|  &  253  &  60  &  0.4  &  224  &  20  &  80.2  &  20  &  0.3  &  20  &  84.7  &  40  &  60.6
 \\

\hline
\verb|oo/autoproof_account|  &  385  &  0  &  0.4  &  446  &  0  &  80.2  &  0  &  0.4  &  0  &  84.8  &  0  &  80.8  \\
\verb|oo/binary_search|  &  68  &  100  &  0.4  &  158  &  67  &  20.1  &  67  &  0.2  &  100  &  0.2  &  100  &  0.2  \\
\verb|oo/joogie_examples|  &  187  &  60  &  0.5  &  277  &  60  &  40.2  &  60  &  0.2  &  60  &  42.4  &  60  &  40.4  \\
\verb|oo/joogie_helloWorld|  &  142  &  50  &  0.5  &  175  &  50  &  20.1  &  50  &  0.1  &  50  &  21.3  &  50  &  20.3  \\
\verb|oo/linked_list_max|  &  44  &  100  &  0.4  &  90  &  100  &  0.2  &  100  &  0.1  &  100  &  0.2  &  100  &  0.1  \\
\verb|oo/rotation_by_copy|  &  52  &  0  &  21.0  &  99  &  0  &  20.1  &  0  &  0.1  &  0  &  21.3  &  100  &  0.3
 \\

\end{tabular}
\caption{Results for the programs in groups \nat (above the horizontal line) and \obj (below it) in the experiments. 
For each program (\textsc{name}) the \Boogie program length in non-comment non-empty lines of code (\textsc{loc}) and the length of its \textsc{\Why} translation; and, for both \Boogie and \Why, for each choice of SMT solver among \textsc{Alt-Ergo}, \textsc{CVC3}, and \textsc{Z3}: the percentage of goals verified in each program (\textsc{\% v.}) and the verification time (\textsc{t}) in seconds (with a timeout of 180 seconds).}
\label{tab:programs-full-boogie-oo}
\end{table}

\clearpage

\begin{scriptsize}
\begin{longtable}{l |r *{1}{rr} |r *{4}{rr}}
\multicolumn{1}{c}{} &
\multicolumn{3}{c}{\textsc{\Boogie}} &
\multicolumn{9}{c}{\textsc{\Why}}
\\
\multicolumn{1}{c}{} &
\multicolumn{1}{c}{} &
\multicolumn{2}{c}{\textsc{Z3}} &
\multicolumn{1}{c}{} &
\multicolumn{2}{c}{\textsc{Alt-Ergo}} &
\multicolumn{2}{c}{\textsc{CVC3}} &
\multicolumn{2}{c}{\textsc{CVC4}} &
\multicolumn{2}{c}{\textsc{Z3}}
\\
\multicolumn{1}{c}{\textsc{name}} &
\multicolumn{1}{c}{\textsc{loc}} &
\multicolumn{1}{c}{\textsc{\% v.}} &
\multicolumn{1}{c}{\textsc{t}} &
\multicolumn{1}{c}{\textsc{loc}} &
\multicolumn{1}{c}{\textsc{\% v.}} &
\multicolumn{1}{c}{\textsc{t}} &
\multicolumn{1}{c}{\textsc{\% v.}} &
\multicolumn{1}{c}{\textsc{t}} &
\multicolumn{1}{c}{\textsc{\% v.}} &
\multicolumn{1}{c}{\textsc{t}} &
\multicolumn{1}{c}{\textsc{\% v.}} &
\multicolumn{1}{c}{\textsc{t}}
\\
\hline
\endhead
\verb|doomed/doomdebug|  &  36  &  0  &  0.3  &  86  &  0  &  40.1  &  0  &  0.1  &  0  &  42.4  &  0  &  40.4  \\
\verb|doomed/doomed|  &  73  &  43  &  0.3  &  185  &  43  &  80.2  &  43  &  0.1  &  43  &  84.7  &  43  &  80.7  \\
\verb|doomed/notdoomed|  &  43  &  50  &  0.3  &  107  &  50  &  40.1  &  50  &  0.1  &  50  &  42.4  &  50  &  40.4  \\
\verb|doomed/smoke0|  &  61  &  67  &  0.3  &  148  &  67  &  40.1  &  67  &  0.2  &  67  &  42.4  &  67  &  40.4  \\
\verb|lock/Lock|  &  86  &  100  &  0.4  &  163  &  67  &  20.1  &  67  &  0.1  &  67  &  21.3  &  67  &  20.2  \\
\verb|lock/LockIncorrect|  &  34  &  0  &  0.3  &  64  &  0  &  20.1  &  0  &  0.1  &  0  &  21.2  &  0  &  20.2  \\
\verb|smoke/smoke0|  &  41  &  100  &  0.3  &  108  &  100  &  0.1  &  100  &  0.1  &  100  &  0.2  &  100  &  0.1  \\
\verb|snapshots/Snapshots0.v0|  &  16  &  0  &  0.3  &  72  &  0  &  80.1  &  0  &  0.1  &  0  &  84.6  &  0  &  80.6  \\
\verb|snapshots/Snapshots0.v1|  &  16  &  50  &  0.3  &  72  &  50  &  40.1  &  50  &  0.1  &  50  &  42.4  &  50  &  40.3  \\
\verb|snapshots/Snapshots0.v2|  &  12  &  67  &  0.3  &  60  &  67  &  20.1  &  67  &  0.1  &  67  &  21.3  &  67  &  20.3  \\
\verb|snapshots/Snapshots1.v0|  &  10  &  50  &  0.3  &  48  &  50  &  20.1  &  50  &  0.1  &  50  &  21.3  &  50  &  20.2  \\
\verb|snapshots/Snapshots1.v1|  &  10  &  50  &  0.3  &  48  &  50  &  20.1  &  50  &  0.1  &  50  &  21.2  &  50  &  20.3  \\
\verb|snapshots/Snapshots1.v2|  &  11  &  50  &  0.3  &  50  &  50  &  20.1  &  50  &  0.1  &  50  &  21.2  &  50  &  20.2  \\
\verb|snapshots/Snapshots10.v0|  &  14  &  100  &  0.3  &  48  &  100  &  0.1  &  100  &  0.1  &  100  &  0.1  &  100  &  0.1  \\
\verb|snapshots/Snapshots10.v1|  &  14  &  100  &  0.3  &  48  &  100  &  0.1  &  100  &  0.1  &  100  &  0.1  &  100  &  0.1  \\
\verb|snapshots/Snapshots11.v0|  &  10  &  0  &  0.3  &  43  &  0  &  20.1  &  0  &  0.1  &  0  &  21.2  &  0  &  20.2  \\
\verb|snapshots/Snapshots11.v1|  &  10  &  0  &  0.3  &  43  &  0  &  20.1  &  0  &  0.1  &  0  &  21.2  &  0  &  20.2  \\
\verb|snapshots/Snapshots12.v0|  &  12  &  100  &  0.4  &  41  &  100  &  0.1  &  100  &  0.1  &  100  &  0.1  &  100  &  0.1  \\
\verb|snapshots/Snapshots12.v1|  &  12  &  0  &  0.3  &  41  &  0  &  20.1  &  0  &  0.1  &  0  &  21.2  &  0  &  20.2  \\
\verb|snapshots/Snapshots13.v0|  &  16  &  100  &  0.3  &  43  &  100  &  0.1  &  100  &  0.0  &  100  &  0.1  &  100  &  0.1  \\
\verb|snapshots/Snapshots13.v1|  &  12  &  0  &  0.4  &  41  &  0  &  20.1  &  0  &  0.1  &  0  &  21.2  &  0  &  20.2  \\
\verb|snapshots/Snapshots14.v0|  &  16  &  100  &  0.3  &  43  &  100  &  0.1  &  100  &  0.1  &  100  &  0.1  &  100  &  0.1  \\
\verb|snapshots/Snapshots14.v1|  &  16  &  0  &  0.3  &  43  &  0  &  20.1  &  0  &  0.1  &  0  &  21.2  &  0  &  20.2  \\
\verb|snapshots/Snapshots15.v0|  &  11  &  100  &  0.3  &  42  &  100  &  0.1  &  100  &  0.1  &  100  &  0.1  &  100  &  0.1  \\
\verb|snapshots/Snapshots15.v1|  &  11  &  0  &  0.3  &  42  &  0  &  20.1  &  0  &  0.1  &  0  &  21.2  &  0  &  20.2  \\
\verb|snapshots/Snapshots16.v0|  &  11  &  100  &  0.3  &  40  &  100  &  0.1  &  0  &  0.1  &  100  &  0.1  &  100  &  0.1  \\
\verb|snapshots/Snapshots16.v1|  &  11  &  0  &  0.3  &  40  &  0  &  20.1  &  0  &  0.1  &  0  &  21.2  &  0  &  20.2  \\
\verb|snapshots/Snapshots17.v0|  &  22  &  100  &  0.3  &  61  &  100  &  0.1  &  100  &  0.1  &  100  &  0.1  &  100  &  0.1  \\
\verb|snapshots/Snapshots17.v1|  &  22  &  0  &  0.3  &  61  &  0  &  20.1  &  0  &  0.1  &  0  &  21.2  &  0  &  20.2  \\
\verb|snapshots/Snapshots18.v0|  &  18  &  100  &  0.3  &  53  &  100  &  0.1  &  100  &  0.1  &  100  &  0.1  &  100  &  0.1  \\
\verb|snapshots/Snapshots18.v1|  &  18  &  0  &  0.3  &  53  &  0  &  20.1  &  0  &  0.1  &  0  &  21.2  &  0  &  20.2  \\
\verb|snapshots/Snapshots19.v0|  &  8  &  0  &  0.3  &  39  &  0  &  20.1  &  0  &  0.1  &  0  &  21.2  &  0  &  20.2  \\
\verb|snapshots/Snapshots19.v1|  &  8  &  0  &  0.3  &  39  &  0  &  20.1  &  0  &  0.1  &  0  &  21.2  &  0  &  20.2  \\
\verb|snapshots/Snapshots2.v0|  &  9  &  100  &  0.3  &  38  &  100  &  0.1  &  100  &  0.1  &  100  &  0.1  &  100  &  0.1  \\
\verb|snapshots/Snapshots2.v1|  &  9  &  100  &  0.3  &  38  &  100  &  0.1  &  100  &  0.1  &  100  &  0.1  &  100  &  0.1  \\
\verb|snapshots/Snapshots2.v2|  &  10  &  100  &  0.3  &  40  &  100  &  0.1  &  100  &  0.1  &  100  &  0.1  &  100  &  0.1  \\
\verb|snapshots/Snapshots2.v3|  &  10  &  100  &  0.3  &  40  &  100  &  0.1  &  100  &  0.1  &  100  &  0.1  &  100  &  0.1  \\
\verb|snapshots/Snapshots2.v4|  &  10  &  100  &  0.3  &  40  &  100  &  0.1  &  100  &  0.1  &  100  &  0.1  &  100  &  0.1  \\
\verb|snapshots/Snapshots2.v5|  &  11  &  100  &  0.3  &  42  &  100  &  0.1  &  100  &  0.1  &  100  &  0.1  &  100  &  0.1  \\
\verb|snapshots/Snapshots20.v0|  &  16  &  0  &  0.3  &  44  &  0  &  20.1  &  0  &  0.1  &  0  &  21.2  &  0  &  20.2  \\
\verb|snapshots/Snapshots20.v1|  &  16  &  0  &  0.3  &  44  &  0  &  20.1  &  0  &  0.1  &  0  &  21.3  &  0  &  20.2  \\
\verb|snapshots/Snapshots21.v0|  &  13  &  0  &  0.3  &  41  &  0  &  20.1  &  0  &  0.1  &  0  &  21.2  &  0  &  20.2  \\
\verb|snapshots/Snapshots21.v1|  &  13  &  0  &  0.3  &  41  &  0  &  20.1  &  0  &  0.1  &  0  &  21.2  &  0  &  20.2  \\
\verb|snapshots/Snapshots22.v0|  &  13  &  0  &  0.3  &  41  &  0  &  20.1  &  0  &  0.1  &  0  &  21.2  &  0  &  20.2  \\
\verb|snapshots/Snapshots22.v1|  &  13  &  100  &  0.3  &  41  &  100  &  0.1  &  100  &  0.1  &  100  &  0.1  &  100  &  0.1  \\
\verb|snapshots/Snapshots23.v0|  &  17  &  50  &  0.3  &  52  &  50  &  20.1  &  50  &  0.1  &  50  &  21.2  &  50  &  20.2  \\
\verb|snapshots/Snapshots23.v1|  &  18  &  50  &  0.3  &  53  &  50  &  20.1  &  50  &  0.1  &  50  &  21.3  &  50  &  20.2  \\
\verb|snapshots/Snapshots23.v2|  &  17  &  50  &  0.3  &  52  &  50  &  20.1  &  50  &  0.1  &  50  &  21.2  &  50  &  20.2  \\
\verb|snapshots/Snapshots24.v0|  &  23  &  0  &  0.3  &  51  &  0  &  20.1  &  0  &  0.1  &  0  &  21.3  &  0  &  20.2  \\
\verb|snapshots/Snapshots24.v1|  &  23  &  0  &  0.3  &  51  &  0  &  20.1  &  0  &  0.1  &  0  &  21.3  &  0  &  20.2  \\
\verb|snapshots/Snapshots25.v0|  &  11  &  0  &  0.3  &  47  &  0  &  20.1  &  0  &  0.1  &  0  &  21.2  &  0  &  20.2  \\
\verb|snapshots/Snapshots25.v1|  &  11  &  0  &  0.3  &  47  &  0  &  20.1  &  0  &  0.1  &  0  &  21.2  &  0  &  20.2  \\
\verb|snapshots/Snapshots26.v0|  &  11  &  0  &  0.3  &  47  &  0  &  20.1  &  0  &  0.1  &  0  &  21.3  &  0  &  20.2  \\
\verb|snapshots/Snapshots26.v1|  &  12  &  0  &  0.3  &  48  &  0  &  20.1  &  0  &  0.1  &  0  &  21.2  &  0  &  20.2  \\
\verb|snapshots/Snapshots27.v0|  &  11  &  0  &  0.3  &  47  &  0  &  20.1  &  0  &  0.1  &  0  &  21.2  &  0  &  20.2  \\
\verb|snapshots/Snapshots27.v1|  &  13  &  0  &  0.3  &  51  &  0  &  20.1  &  0  &  0.1  &  0  &  21.2  &  0  &  20.2  \\
\verb|snapshots/Snapshots28.v0|  &  11  &  100  &  0.3  &  48  &  100  &  0.1  &  100  &  0.1  &  100  &  0.1  &  100  &  0.1  \\
\verb|snapshots/Snapshots28.v1|  &  12  &  0  &  0.4  &  48  &  0  &  20.1  &  0  &  0.1  &  0  &  21.2  &  0  &  20.2  \\
\verb|snapshots/Snapshots29.v0|  &  11  &  100  &  0.3  &  47  &  0  &  20.1  &  0  &  0.1  &  0  &  21.2  &  0  &  20.2  \\
\verb|snapshots/Snapshots29.v1|  &  11  &  0  &  0.3  &  47  &  0  &  20.1  &  0  &  0.1  &  0  &  21.2  &  0  &  20.2  \\
\verb|snapshots/Snapshots3.v0|  &  13  &  100  &  0.3  &  41  &  100  &  0.1  &  100  &  0.1  &  100  &  0.1  &  100  &  0.1  \\
\verb|snapshots/Snapshots3.v1|  &  13  &  0  &  0.3  &  41  &  0  &  20.1  &  0  &  0.1  &  0  &  21.2  &  0  &  20.2  \\
\verb|snapshots/Snapshots30.v0|  &  11  &  0  &  0.3  &  42  &  0  &  20.1  &  0  &  0.1  &  0  &  21.2  &  0  &  20.2  \\
\verb|snapshots/Snapshots30.v1|  &  12  &  0  &  0.3  &  43  &  0  &  20.1  &  0  &  0.1  &  0  &  21.2  &  0  &  20.2  \\
\verb|snapshots/Snapshots31.v0|  &  12  &  100  &  0.3  &  44  &  100  &  0.1  &  100  &  0.1  &  100  &  0.1  &  100  &  0.1  \\
\verb|snapshots/Snapshots31.v1|  &  11  &  0  &  0.4  &  43  &  0  &  20.1  &  0  &  0.1  &  0  &  21.2  &  0  &  20.2  \\
\verb|snapshots/Snapshots32.v0|  &  12  &  100  &  0.4  &  44  &  100  &  0.1  &  100  &  0.1  &  100  &  0.1  &  100  &  0.1  \\
\verb|snapshots/Snapshots32.v1|  &  9  &  0  &  0.3  &  41  &  0  &  20.1  &  0  &  0.1  &  0  &  21.2  &  0  &  20.2  \\
\verb|snapshots/Snapshots33.v0|  &  12  &  100  &  0.3  &  44  &  100  &  0.1  &  100  &  0.1  &  100  &  0.1  &  100  &  0.1  \\
\verb|snapshots/Snapshots33.v1|  &  6  &  100  &  0.4  &  37  &  100  &  0.1  &  100  &  0.1  &  100  &  0.1  &  100  &  0.1  \\
\verb|snapshots/Snapshots34.v0|  &  6  &  100  &  0.3  &  38  &  100  &  0.1  &  100  &  0.1  &  100  &  0.1  &  100  &  0.1  \\
\verb|snapshots/Snapshots34.v1|  &  5  &  0  &  0.3  &  36  &  0  &  20.1  &  0  &  0.1  &  0  &  21.2  &  0  &  20.2  \\
\verb|snapshots/Snapshots35.v0|  &  6  &  100  &  0.3  &  38  &  100  &  0.1  &  100  &  0.1  &  100  &  0.1  &  100  &  0.1  \\
\verb|snapshots/Snapshots35.v1|  &  5  &  0  &  0.3  &  36  &  0  &  20.1  &  0  &  0.1  &  0  &  21.2  &  0  &  20.2  \\
\verb|snapshots/Snapshots36.v0|  &  11  &  100  &  0.4  &  44  &  100  &  0.1  &  0  &  0.1  &  100  &  0.1  &  100  &  0.1  \\
\verb|snapshots/Snapshots36.v1|  &  11  &  0  &  0.3  &  44  &  0  &  20.1  &  0  &  0.1  &  0  &  21.2  &  0  &  20.2  \\
\verb|snapshots/Snapshots37.v0|  &  7  &  100  &  0.3  &  42  &  100  &  0.1  &  0  &  0.1  &  100  &  0.1  &  100  &  0.1  \\
\verb|snapshots/Snapshots37.v1|  &  7  &  0  &  0.3  &  42  &  0  &  20.1  &  0  &  0.1  &  0  &  21.2  &  0  &  20.2  \\
\verb|snapshots/Snapshots38.v0|  &  10  &  100  &  0.4  &  43  &  100  &  0.1  &  100  &  0.1  &  100  &  0.1  &  100  &  0.1  \\
\verb|snapshots/Snapshots38.v1|  &  11  &  0  &  0.3  &  44  &  0  &  20.1  &  0  &  0.1  &  0  &  21.2  &  0  &  20.2  \\
\verb|snapshots/Snapshots38.v2|  &  11  &  100  &  0.3  &  44  &  100  &  0.1  &  100  &  0.1  &  100  &  0.1  &  100  &  0.1  \\
\verb|snapshots/Snapshots39.v0|  &  10  &  100  &  0.3  &  43  &  100  &  0.1  &  100  &  0.1  &  100  &  0.1  &  100  &  0.1  \\
\verb|snapshots/Snapshots39.v1|  &  11  &  0  &  0.4  &  44  &  0  &  20.1  &  0  &  0.1  &  0  &  21.2  &  0  &  20.2  \\
\verb|snapshots/Snapshots39.v2|  &  11  &  100  &  0.3  &  44  &  100  &  0.1  &  0  &  0.1  &  100  &  0.1  &  100  &  0.1  \\
\verb|snapshots/Snapshots4.v0|  &  23  &  100  &  0.3  &  64  &  100  &  0.1  &  100  &  0.1  &  100  &  0.1  &  100  &  0.2  \\
\verb|snapshots/Snapshots4.v1|  &  27  &  50  &  0.3  &  76  &  50  &  40.1  &  50  &  0.1  &  50  &  42.4  &  50  &  40.4  \\
\verb|snapshots/Snapshots40.v0|  &  11  &  0  &  0.3  &  44  &  0  &  20.1  &  0  &  0.1  &  0  &  21.2  &  0  &  20.2  \\
\verb|snapshots/Snapshots40.v1|  &  12  &  0  &  0.3  &  45  &  0  &  20.1  &  0  &  0.1  &  0  &  21.2  &  0  &  20.2  \\
\verb|snapshots/Snapshots40.v2|  &  12  &  0  &  0.3  &  45  &  0  &  20.0  &  0  &  0.1  &  0  &  21.2  &  0  &  20.2  \\
\verb|snapshots/Snapshots41.v0|  &  31  &  40  &  0.3  &  99  &  40  &  60.2  &  40  &  0.1  &  40  &  63.5  &  40  &  60.5  \\
\verb|snapshots/Snapshots41.v1|  &  31  &  40  &  0.3  &  100  &  40  &  60.1  &  40  &  0.2  &  40  &  63.5  &  40  &  60.5  \\
\verb|snapshots/Snapshots5.v0|  &  9  &  100  &  0.3  &  39  &  100  &  0.1  &  100  &  0.1  &  100  &  0.1  &  100  &  0.1  \\
\verb|snapshots/Snapshots5.v1|  &  9  &  0  &  0.3  &  39  &  0  &  20.1  &  0  &  0.1  &  0  &  21.2  &  0  &  20.2  \\
\verb|snapshots/Snapshots6.v0|  &  12  &  100  &  0.3  &  43  &  100  &  0.1  &  100  &  0.1  &  100  &  0.1  &  100  &  0.1  \\
\verb|snapshots/Snapshots6.v1|  &  12  &  0  &  0.3  &  43  &  0  &  20.1  &  0  &  0.1  &  0  &  21.2  &  0  &  20.2  \\
\verb|snapshots/Snapshots7.v0|  &  14  &  100  &  0.3  &  45  &  100  &  0.1  &  100  &  0.1  &  100  &  0.1  &  100  &  0.1  \\
\verb|snapshots/Snapshots7.v1|  &  14  &  100  &  0.3  &  45  &  100  &  0.1  &  100  &  0.1  &  100  &  0.1  &  100  &  0.1  \\
\verb|snapshots/Snapshots8.v0|  &  11  &  100  &  0.3  &  45  &  100  &  0.1  &  100  &  0.1  &  100  &  0.1  &  100  &  0.1  \\
\verb|snapshots/Snapshots8.v1|  &  11  &  100  &  0.3  &  45  &  100  &  0.1  &  100  &  0.1  &  100  &  0.1  &  100  &  0.1  \\
\verb|snapshots/Snapshots9.v0|  &  13  &  100  &  0.4  &  47  &  100  &  0.1  &  100  &  0.1  &  100  &  0.1  &  100  &  0.1  \\
\verb|snapshots/Snapshots9.v1|  &  11  &  100  &  0.3  &  45  &  100  &  0.1  &  100  &  0.1  &  100  &  0.1  &  100  &  0.1  \\
\verb|test13/ErrorTraceTestLoopInvViolationBPL|  &  19  &  0  &  0.3  &  86  &  0  &  60.1  &  0  &  0.1  &  0  &  63.5  &  0  &  60.5  \\
\verb|test15/CaptureState|  &  23  &  0  &  0.3  &  62  &  0  &  20.1  &  0  &  0.1  &  0  &  21.2  &  0  &  20.2  \\
\verb|test15/InterpretedFunctionTests|  &  15  &  0  &  0.3  &  66  &  0  &  60.1  &  0  &  0.1  &  0  &  63.5  &  0  &  60.5  \\
\verb|test15/IntInModel|  &  3  &  0  &  0.3  &  36  &  0  &  20.1  &  0  &  0.1  &  0  &  21.2  &  0  &  20.2  \\
\verb|test15/ModelTest|  &  10  &  0  &  0.3  &  49  &  0  &  20.1  &  0  &  0.1  &  0  &  21.2  &  0  &  20.2  \\
\verb|test15/NullInModel|  &  5  &  0  &  0.3  &  39  &  0  &  20.1  &  0  &  0.1  &  0  &  21.2  &  0  &  20.2  \\
\verb|test16/LoopUnroll|  &  63  &  0  &  0.3  &  124  &  0  &  60.1  &  0  &  0.1  &  0  &  63.5  &  0  &  60.5  \\
\verb|test17/contractinfer|  &  21  &  0  &  0.3  &  68  &  0  &  40.1  &  0  &  0.1  &  0  &  42.4  &  0  &  40.3  \\
\verb|test2/AssertVerifiedUnder0|  &  26  &  50  &  0.3  &  100  &  0  &  120.2  &  0  &  0.2  &  0  &  126.9  &  0  &  120.9  \\
\verb|test2/AssumeEnsures|  &  53  &  57  &  0.3  &  124  &  57  &  60.2  &  57  &  0.2  &  57  &  63.6  &  57  &  60.5  \\
\verb|test2/AssumptionVariables0|  &  44  &  50  &  0.3  &  137  &  0  &  120.2  &  0  &  0.1  &  0  &  126.9  &  0  &  120.9  \\
\verb|test2/Axioms|  &  24  &  67  &  0.3  &  73  &  67  &  20.1  &  67  &  0.1  &  67  &  21.3  &  67  &  20.2  \\
\verb|test2/B|  &  65  &  100  &  0.3  &  112  &  0  &  80.1  &  0  &  0.1  &  0  &  84.6  &  0  &  80.7  \\
\verb|test2/Call|  &  49  &  40  &  0.3  &  117  &  20  &  80.2  &  20  &  0.1  &  20  &  84.6  &  20  &  80.6  \\
\verb|test2/ContractEvaluationOrder|  &  26  &  25  &  0.3  &  101  &  25  &  60.1  &  25  &  0.1  &  25  &  63.5  &  25  &  60.5  \\
\verb|test2/CutBackEdge|  &  35  &  20  &  0.3  &  96  &  0  &  100.2  &  0  &  0.1  &  0  &  105.8  &  0  &  100.7  \\
\verb|test2/Ensures|  &  61  &  50  &  0.5  &  168  &  50  &  100.2  &  50  &  0.3  &  50  &  105.9  &  50  &  100.8  \\
\verb|test2/False|  &  14  &  100  &  0.3  &  54  &  100  &  0.1  &  100  &  0.1  &  100  &  0.1  &  100  &  0.1  \\
\verb|test2/FormulaTerm2|  &  36  &  50  &  0.3  &  104  &  50  &  40.1  &  50  &  0.2  &  50  &  42.4  &  50  &  40.4  \\
\verb|test2/FreeCall|  &  59  &  64  &  0.5  &  185  &  27  &  160.2  &  27  &  0.2  &  27  &  169.2  &  27  &  161.2  \\
\verb|test2/Implies|  &  28  &  0  &  0.3  &  97  &  0  &  100.1  &  0  &  0.1  &  0  &  105.7  &  0  &  100.7  \\
\verb|test2/InvariantVerifiedUnder0|  &  42  &  17  &  0.3  &  146  &  0  &  120.2  &  0  &  0.2  &  0  &  126.9  &  0  &  120.9  \\
\verb|test2/LoopInvAssume|  &  15  &  0  &  0.3  &  44  &  0  &  20.1  &  0  &  0.1  &  0  &  21.2  &  0  &  20.2  \\
\verb|test2/Passification|  &  155  &  64  &  0.5  &  290  &  18  &  180.0  &  18  &  0.3  &  18  &  180.0  &  18  &  180.0  \\
\verb|test2/Quantifiers|  &  122  &  57  &  0.4  &  254  &  86  &  40.3  &  64  &  0.3  &  79  &  63.8  &  93  &  20.3  \\
\verb|test2/SelectiveChecking|  &  31  &  25  &  0.3  &  121  &  0  &  80.1  &  0  &  0.1  &  0  &  84.6  &  0  &  80.6  \\
\verb|test2/sk_hack|  &  17  &  100  &  0.3  &  44  &  0  &  20.1  &  0  &  0.1  &  0  &  21.2  &  0  &  20.2  \\
\verb|test2/Timeouts0|  &  71  &  0  &  1.3  &  156  &  0  &  60.1  &  0  &  0.2  &  0  &  63.5  &  0  &  60.5  \\
\verb|test2/TypeEncodingM|  &  19  &  0  &  0.3  &  60  &  0  &  20.1  &  0  &  0.1  &  0  &  21.2  &  0  &  20.2  \\
\verb|test21/BooleanQuantification2|  &  9  &  0  &  0.3  &  46  &  0  &  20.1  &  0  &  0.0  &  0  &  21.2  &  0  &  20.2  \\
\verb|test21/Boxing|  &  15  &  0  &  0.5  &  49  &  0  &  20.1  &  0  &  0.1  &  0  &  21.2  &  0  &  20.2  \\
\verb|test21/Casts|  &  7  &  0  &  0.3  &  48  &  0  &  20.1  &  0  &  0.1  &  0  &  21.2  &  0  &  20.2  \\
\verb|test21/Colors|  &  13  &  0  &  0.3  &  62  &  0  &  40.1  &  0  &  0.1  &  0  &  42.4  &  0  &  40.4  \\
\verb|test21/DisjointDomains|  &  21  &  0  &  0.4  &  81  &  0  &  60.1  &  0  &  0.2  &  0  &  63.5  &  0  &  60.5  \\
\verb|test21/EmptySetBug|  &  18  &  0  &  0.4  &  54  &  0  &  20.1  &  0  &  0.1  &  0  &  21.2  &  0  &  20.2  \\
\verb|test21/FunAxioms|  &  24  &  50  &  0.5  &  81  &  50  &  20.1  &  50  &  0.1  &  50  &  21.3  &  50  &  20.2  \\
\verb|test21/FunAxioms2|  &  13  &  0  &  0.3  &  49  &  0  &  20.1  &  0  &  0.1  &  0  &  21.2  &  0  &  20.2  \\
\verb|test21/InterestingExamples3|  &  17  &  67  &  0.4  &  71  &  33  &  40.1  &  33  &  0.2  &  33  &  42.4  &  33  &  40.4  \\
\verb|test21/InterestingExamples5|  &  9  &  100  &  0.3  &  45  &  100  &  0.1  &  0  &  0.1  &  100  &  0.1  &  100  &  0.1  \\
\verb|test21/Keywords|  &  5  &  100  &  0.3  &  38  &  100  &  0.1  &  100  &  0.1  &  100  &  0.1  &  100  &  0.1  \\
\verb|test21/LargeLiterals0|  &  12  &  0  &  0.3  &  46  &  0  &  20.1  &  0  &  0.1  &  0  &  21.2  &  0  &  20.2  \\
\verb|test21/LetSorting|  &  11  &  100  &  0.3  &  43  &  0  &  20.1  &  0  &  0.1  &  0  &  21.2  &  0  &  20.2  \\
\verb|test21/Maps2|  &  14  &  100  &  0.3  &  52  &  100  &  0.1  &  0  &  0.1  &  100  &  0.2  &  0  &  20.2  \\
\verb|test21/Orderings|  &  13  &  50  &  0.4  &  59  &  0  &  40.1  &  0  &  0.1  &  0  &  42.4  &  0  &  40.4  \\
\verb|test21/Orderings2|  &  11  &  0  &  0.4  &  49  &  0  &  20.1  &  0  &  0.1  &  0  &  21.2  &  0  &  20.2  \\
\verb|test21/Orderings3|  &  22  &  0  &  0.4  &  78  &  0  &  40.1  &  0  &  0.1  &  0  &  42.4  &  0  &  40.4  \\
\verb|test21/Orderings4|  &  7  &  0  &  0.4  &  47  &  0  &  20.1  &  0  &  0.1  &  0  &  21.2  &  0  &  20.2  \\
\verb|test21/PolyList|  &  35  &  0  &  0.4  &  91  &  0  &  40.1  &  0  &  0.2  &  0  &  42.4  &  0  &  40.4  \\
\verb|test21/Triggers0|  &  34  &  50  &  0.4  &  92  &  50  &  20.1  &  50  &  0.1  &  0  &  42.4  &  50  &  20.2  \\
\verb|test21/Triggers1|  &  12  &  0  &  0.4  &  50  &  0  &  20.1  &  0  &  0.1  &  0  &  21.2  &  0  &  20.2  \\
\verb|test7/MultipleErrors|  &  14  &  0  &  0.3  &  42  &  0  &  20.1  &  0  &  0.1  &  0  &  21.2  &  0  &  20.2  \\
\verb|test7/NestedVC|  &  20  &  50  &  0.3  &  61  &  0  &  40.1  &  0  &  0.1  &  0  &  42.4  &  0  &  40.3  \\
\verb|test7/UnreachableBlocks|  &  34  &  100  &  0.3  &  79  &  50  &  40.1  &  50  &  0.1  &  50  &  42.4  &  50  &  40.4  \\
\verb|textbook/Bubble|  &  47  &  100  &  0.4  &  110  &  0  &  20.1  &  0  &  0.1  &  0  &  21.3  &  0  &  20.2  \\
\verb|textbook/DutchFlag|  &  47  &  100  &  0.3  &  92  &  0  &  20.1  &  0  &  0.1  &  0  &  21.2  &  0  &  20.2  \\
\verb|textbook/Find|  &  27  &  100  &  0.3  &  72  &  50  &  20.1  &  50  &  0.1  &  50  &  21.3  &  50  &  20.3  \\
\verb|textbook/McCarthy-91|  &  11  &  100  &  0.3  &  47  &  100  &  0.1  &  100  &  0.1  &  100  &  0.1  &  100  &  0.1  \\
\verb|textbook/TuringFactorial|  &  27  &  100  &  0.3  &  81  &  0  &  20.1  &  0  &  0.1  &  0  &  21.2  &  0  &  20.2
 \\

\caption{Results for the programs in group \tes in the experiments.
The measures are the same as in \autoref{tab:programs-full-boogie-oo}.}
\label{tab:programs-full-tests}
\end{longtable}
\end{scriptsize}

\fi

\end{document}